\documentclass[a4paper,12pt]{article}
\usepackage[margin=1in]{geometry}
\usepackage{setspace}
\usepackage{float}
\usepackage{epstopdf}
\usepackage{standalone}
\usepackage{subfig}
\usepackage{amsmath}
\usepackage{pdflscape}
\usepackage{authblk}
\usepackage{graphicx}
\usepackage[flushleft]{threeparttable}
\usepackage[normalem]{ulem}
\usepackage{soul}
\usepackage[comma,authoryear]{natbib}
\usepackage[citecolor=blue]{hyperref}

\title{Inferring Ice Thickness from a Glacier Dynamics Model and Multiple Surface Datasets}
\author[1]{Yawen Guan}
\author[1]{Murali Haran\thanks{Corresponding author: Prof. M. Haran, Department of Statistics, The Pennsylvania State University, University Park, PA 16802, USA. Email: muh10@psu.edu}}
\author[2]{David Pollard}
\affil[1]{\footnotesize{Department of Statistics, The Pennsylvania State University, University Park, USA}}
\affil[2]{\footnotesize{Earth and Environmental Systems Institude, The Pennsylvania State University, USA}}
\newcommand{\bs}[1]{\boldsymbol{#1}}
\begin{document}
	\maketitle
	
	\begin{abstract}
		The future behavior of the West Antarctic Ice Sheet (WAIS) may have a major impact on future climate. For instance, ice sheet melt may contribute significantly to global sea level rise. Understanding the current state of WAIS is therefore of great interest. WAIS is drained by fast-flowing glaciers which are major contributors to ice loss. Hence, understanding the stability and dynamics of glaciers is critical for predicting the future of the ice sheet. Glacier dynamics are driven by the interplay between the topography, temperature and basal conditions beneath the ice. A glacier dynamics model describes the interactions between these processes. We develop a hierarchical Bayesian model that integrates multiple ice sheet surface data sets with a glacier dynamics model. Our approach allows us to (1) infer important parameters describing the glacier dynamics, (2) learn about ice sheet thickness, and (3) account for errors in the observations and the model. Because we have relatively dense and accurate ice thickness data from the Thwaites Glacier in West Antarctica, we use these data to validate the proposed approach. The long-term goal of this work is to  have a general model that may be used to study multiple glaciers in the Antarctic.
		\textbf{Keywords}: ice sheet, glacier dynamics, hierarchical Bayes, Gaussian process, Markov chain Monte Carlo, West Antarctic ice sheet. 
	\end{abstract}
	
	
	\section{Introduction}
	
	A common challenge in glaciology is the estimation of unknown bed properties below glaciers and ice sheets using high-quality surface observations and some knowledge or models of the glacial physics. Here we focus on estimating or interpolating poorly known bedrock topography (or equivalently ice thickness) based on well understood ice sheet physics and higher quality information regarding surface quantities. The principal physics commonly applied is the conservation of ice mass; given high-quality observations of surface elevation, ice velocity and surface mass balance, we can then deduce the ice thickness assuming the ice sheet is in a steady state.  Attempts based on this principle have been made over continental Antarctica with very coarse grid spacing \citep{warner2000} and for a major glacier in northeast and in other locations of Greenland using a much higher-resolution spacing \citep{morlighem2011mass,morlighem2013high,morlighem2014deeply}. For simplicity, we focus on a 1-D flowline along the centerline of a glacier with two improvements to the physical model: (1) we add a new component to the dynamics model, the Shallow-Ice Approximation (SIA), which goes beyond mass conservation for shearing flow, and (2) we include the varying glacier width to account for tributaries, which contribute to mass flux in the downstream.

	A purely statistical interpolation approach for deriving ice thickness will violate the underlying physics, resulting in a physically implausible reconstruction of the ice sheet. Furthermore, it would not contribute to our understanding of ice sheet dynamics which is intrinsically of scientific interest. On the other hand, a purely physics-based approach using the simple physical model may not capture the entire glacier process no matter how well we adjust the model parameters, because it ignores various errors and uncertainties resulting in estimates that are over-confident and unreliable. Therefore, we need a modeling approach that combines both the physics as well as the observational data sets, and allows for errors and data-model discrepancies. We propose an approach that allows us to: (1) combine multiple data sources and simple physical laws for deducing ice thickness, (2) estimate the key parameter of the physical model along with their uncertainties, and (3) account for errors and for data-model discrepancies. Our main focus is on parameter estimation of the physical model and reconstructing the ice thickness using sparsely observed thickness data, high-quality surface observations and well-understood ice sheet physics.

	Efforts have been made on modeling glacier using a Bayesian approach with a focus on smoothing resolutions for the North East Ice Stream in Greenland \citep[cf.][]{Berliner2008,berliner2008modeling}. A mass conservation approach, similar to the one we use here, was employed by \cite{Berliner2008,berliner2008modeling} for a 1-dimensional glacier dynamics model. This was used to relate basal (bed) elevation to surface velocity speed and surface elevation. In their study, they use all available ice thickness, the difference between the surface and basal elevations to estimate the ice deformation coefficient and to select models of different smoothing resolutions. The problem we consider in this paper differs from their work by posing at least two significant additional challenges: (1) improvements to the glacier dynamics model require us to estimate the flow width, which is an unknown quantity that varies along the flowline, and (2) we use sparse thickness observations to reconstruct the ice profile. Problems that involve combining a physical model and observations for improving estimation and prediction have been an active research area. Some studies focus on combining multiple information from deterministic physical model outputs and observations for improving estimation of a spatial temporal field without explicitly solving the physical model \citep[cf.][]{ReichFuentes2007, WikleJASA2001}; because running the physical model is computationally intensive. The glacier dynamics model we propose has a simple form that can be solved explicitly and efficiently, allowing us to work with the model directly. Others solve the physical model explicitly with a focus on inferring model parameters, which are
	typically scalars, from noisy observations \citep[cf.][]{Li2005, RamsayODE, Xun2013}. However, it is not clear how to adapt these approaches to our problem.  
	Our problem and methodology is different in the following ways. (1) Our variable of main interest, ice thickness, is highly nonlinearly related to other physical model inputs. Therefore, a basis function approximation to ice thickness using generalized smoothing approaches \citep{RamsayODE, Xun2013} does not simplify the form of the differential equation; the estimation problem in a statistical formulation therefore remains difficult. (2) The unknown flow width parameter is spatially varying. Therefore, it is impossible to approximate the Hessian matrix required by optimization approaches \citep[cf.][]{Li2005}. (3) Moreover, the multiple data inputs to the physical model have different spatial resolutions and are subject to observational errors. These challenges motivate the Bayesian hierarchical approach proposed in Section \ref{sec:3}.

	We apply our method to the Thwaites Glacier in West Antarctica, because (1) a relatively high-quality dataset of all necessary surface observations is available along its centerline (see Section \ref{sec:data}), (2) Thwaites Glacier is the center of considerable current attention and concern, as it may be the first major Antarctic outlet to undergo dramatic future retreat \citep{rignot2014, joughin2014, deconto2016}, and (3) it is not in steady state, but has experienced rapid thinning in recent decades \citep{pritchard2012, rignot2014}, and hence provides a good testbed for our method.

	The outline of the remainder of the paper is as follows. We start by introducing in Section \ref{sec:2} the physical model and the available data sets. We then introduce the statistical model and its inference in Section \ref{sec:3}. We present the results of the Bayesian approach on simulated data sets and application to the Thwaites Glacier in Section \ref{sec:4}. We discuss the extension of the proposed method to other glaciers in Section \ref{sec:5}.

	\section{Glacier dynamics and data}\label{sec:2}
	\subsection{Glacier dynamics}
\begin{figure}[!htbp]
		\centering
		\subfloat[The red, green, and blue dashed lines represent three pairs of streamlines, and the centerline (white dashed) for the Thwaites Glacier.]{\includegraphics[width=0.7\textwidth]{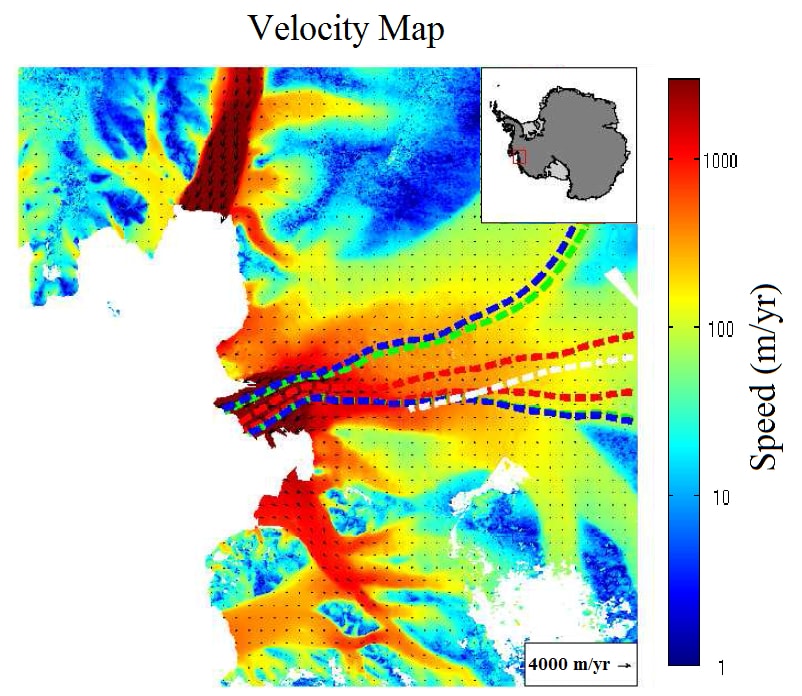}{\label{fig:flowlinea}}}
		\vspace{0mm}
		\subfloat[Flow widths obtained from taking the transverse distance perpendicular to the centerline.]{\includegraphics[width=0.7\textwidth]{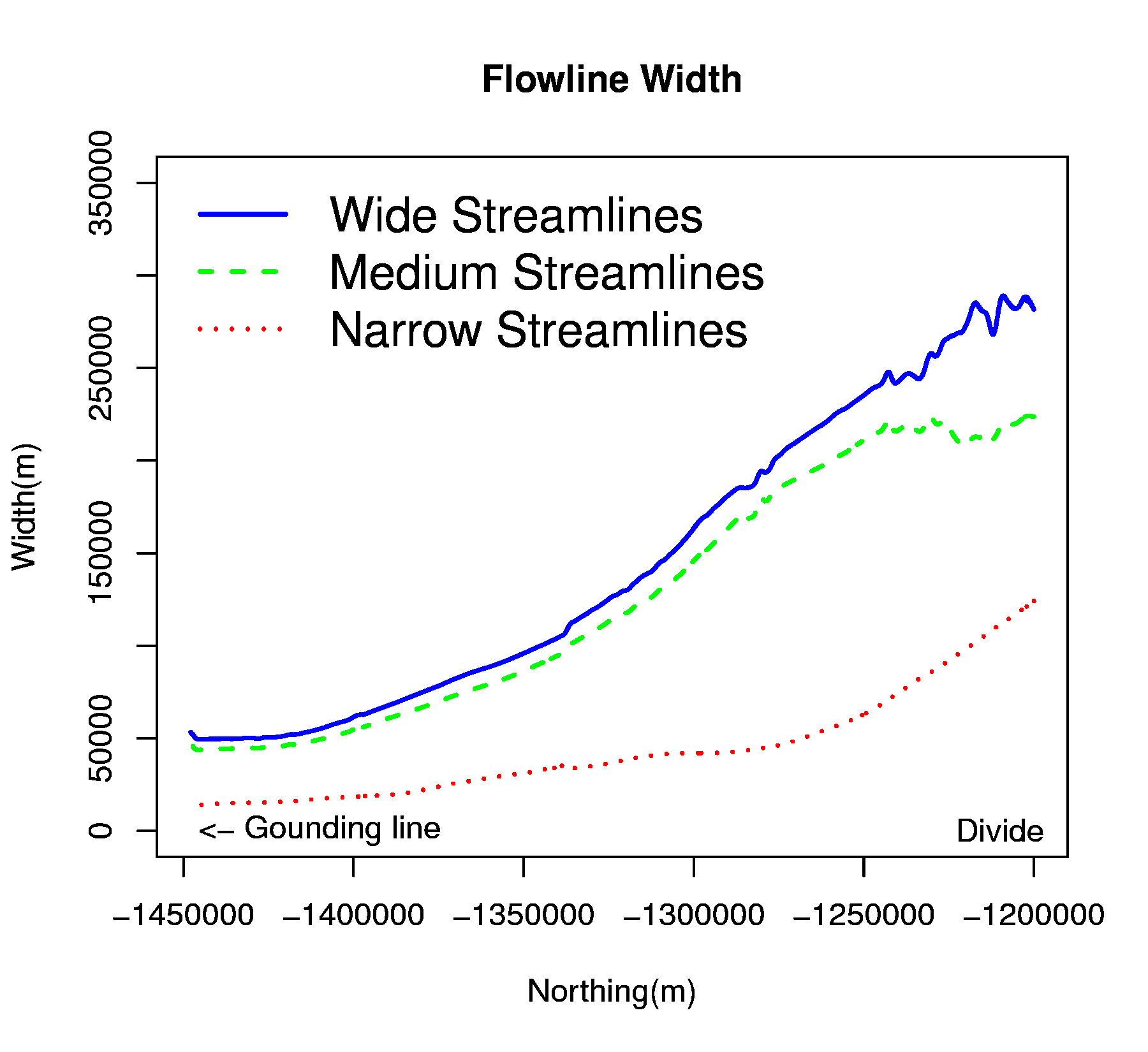}{\label{fig:flowlineb}}}
		\caption{Depending on different starting locations at the divide, the flow widths vary.}
		\label{fig:flowline}
	\end{figure}

	The physical model we employ is based on conservation of ice mass along a flowline, bounded by streamlines separated by the width of the glacier, with an additional physics component, the Shallow-Ice Approximation (SIA), for shearing flow. The spatial domain of the model is 1-dimensional along the centerline, which represents the fastest flow path of a glacier, bounded by the divide (initial location) and upstream of the grounding line (Figure \ref{fig:flowlinea}). The physical model relates several processes, for instance, surface elevation, surface velocity of ice flow, and surface accumulation and thinning rates to ice thickness, allowing us to combine multiple data sets in a physically plausible manner.

	The principal physics used is conservation of ice mass. This implies the change in mass flux downstream is equal to the net accumulation rate and thinning rate. That is, the glacial flow of ice out of a control volume in the downstream direction, minus the flow into the volume from upstream, must equal the net rate of ice externally added on the surface (snowfall minus melt, with no melt occurring here), plus any observed reduction in ice thickness (the thinning rate),
	\begin{equation}\label{eqn:sia}
	\frac{\partial}{\partial x} \left(  \bar{v}(x)h(x)\omega(x) \right) = \left( a(x)- {\tau}(x)\right) \omega(x),
	\end{equation}
	where $ x $ indicates location along the flowline (in meters), $ \bar{v}(x)  $ is the centerline depth-averaged and width-averaged velocity (m/year), $ h(x) $ represents the width-average thickness (m), $ \omega(x) $ is the flow width to account for downstream variations in glacier width, $ a(x) $ is local surface net accumulation rate (m/year), and $ {\tau} = \frac{\partial h(x)}{\partial t}  $ is the ice-column thickness thinning rate (m/year). This equation states that the mass is balanced between the change in flux  along downstream and the net change in the centerline depth, the left and right side of Equation \eqref{eqn:sia} respectively.

	The glacier width $ \omega(x) $ does not necessarily represent the extreme edges of the glacier; it is the transverse distance between a pair of streamlines across which no ice flows out of the domain. If the flowline extends upstream beyond where the glacier splits into tributaries, in principle the glacier width includes the tributaries too, because they contribute to mass flux into the downstream portions of the domain. This is the case for the Thwaites application (see Figure \ref{fig:flowline}). It is assumed that the observed centerline velocities are close to uniform across the width of the glacier; a correction factor could be applied in follow-up work to account for side drag and lateral shear.

	This approach closely follows \cite{morlighem2011mass}, who used
	conservation of mass to deduce ice thicknesses on a major Greenland outlet glacier. They used observed surface velocity $ v_s(x) $ for depth-averaged velocity $ \bar{v}(x) $. Here, we apply a correction to $ v_s(x) $ based on the Shallow Ice Approximation \citep[e.g.][]{van2013fundamentals}.
	to account for the internal ice deformation due to gravitational stress and basal drag due to friction at the ice bed. Let $ v_b(x) $ denote the basal sliding velocity, $ v_s^{SIA}(x) $ and $ \bar{v}^{SIA}(x) $ denote SIA-modeled surface velocity and depth averaged velocity respectively. Our correction to  surface velocity is
	\begin{align}\nonumber
	\bar{v}(x) & = v_s(x) - (v_s^{SIA}(x) - \bar{v}^{SIA}(x)),
	\end{align} where
	\begin{align}\nonumber v_s^{SIA}(x) &= \frac{A}{4} \left( \rho g s(x) \right) ^3  h(x)^4+ v_b(x), \\ \nonumber
	\bar{v}^{SIA}(x) &= \frac{A}{5} \left( \rho g s(x) \right) ^3  h(x)^4+ v_b(x),
	\end{align} so that
	\begin{align}\label{eqn:velcorrection}
	\bar{v}(x) &= v_s(x) - \frac{A}{20} \left( \rho g s(x) \right) ^3 h(x)^4.
	\end{align}
	$ A $ is the ice rheologic coefficient ($ Pa^{-3}/year$) in Glen’s Law with exponent 3, $\rho $ is ice density ($ kg m^{-3} $), $ g $ is gravitational acceleration ($ m s^{-2} $), and $ s(x) $ is the observed downstream surface slope derived from surface elevation. A scalar value of the rheological coefficient $ A $ is used, neglecting its dependence on ice temperature or fabric variations along the centerline. Note that the difference between $ v_s^{SIA}(x) $ and $ \bar{v}^{SIA}(x) $ is used as a correction to the observed velocity, so that the unknown basal velocity $ v_b(x) $ cancels. Thus, no explicit knowledge of basal sliding (and the effect of frozen versus thawed beds) is needed, because its contribution to the depth-averaged transport is included implicitly via the observed surface velocity.
	
\begin{figure}[!htbp]
		\centering
		\captionsetup[subfigure]{width=0.4\textwidth}
		\subfloat[Solutions corresponding to the widest flowline width (blue line in Figure \ref{fig:flowlineb})]{\includegraphics[width=0.5\textwidth,clip, trim=0cm 0cm 0cm 1.2cm]{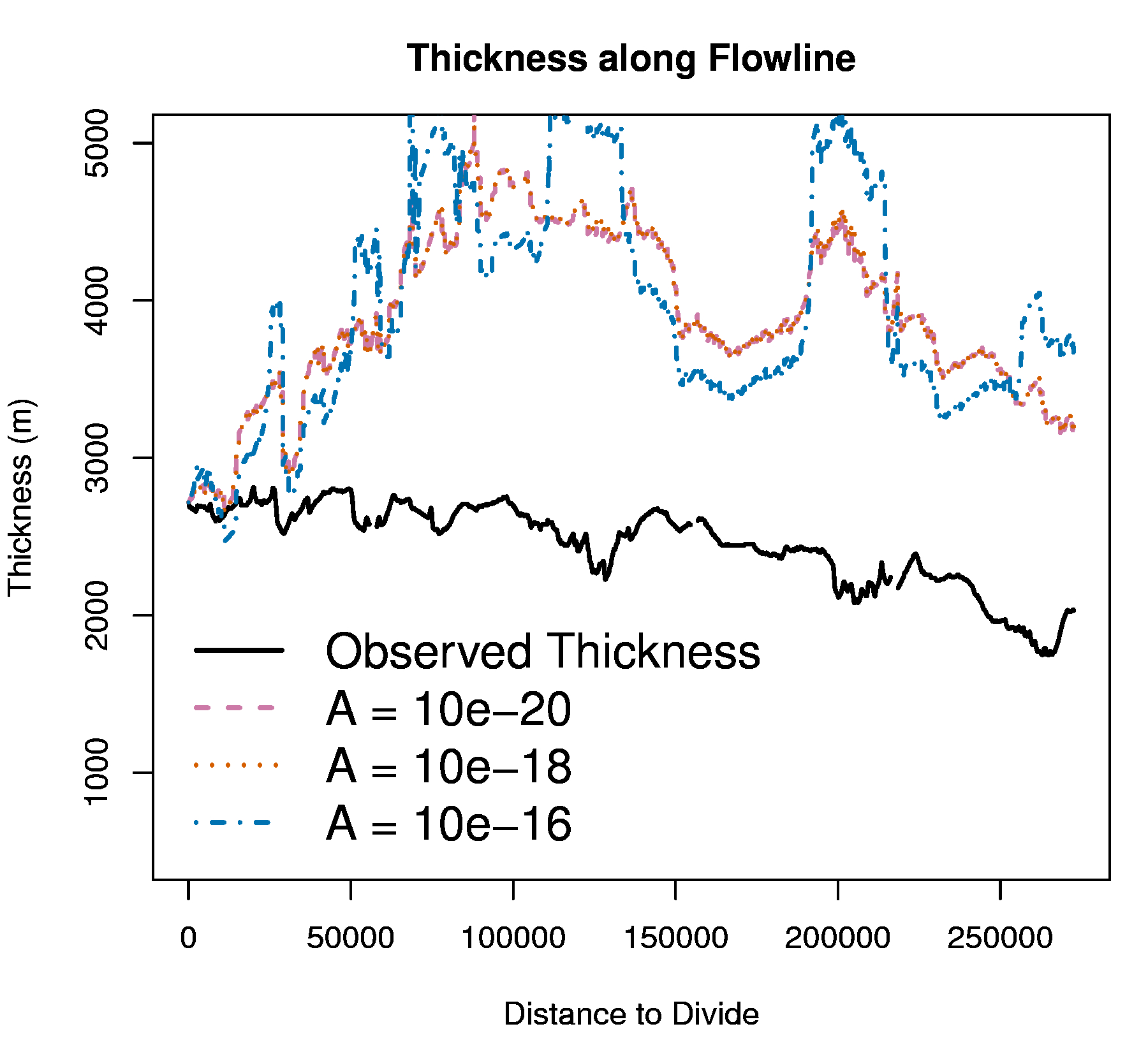}}
		\subfloat[Solutions corresponding to the medium wide flowline width (green line in Figure \ref{fig:flowlineb})]{\includegraphics[width=0.5\textwidth,clip, trim=0cm 0cm 0cm 1.2cm]{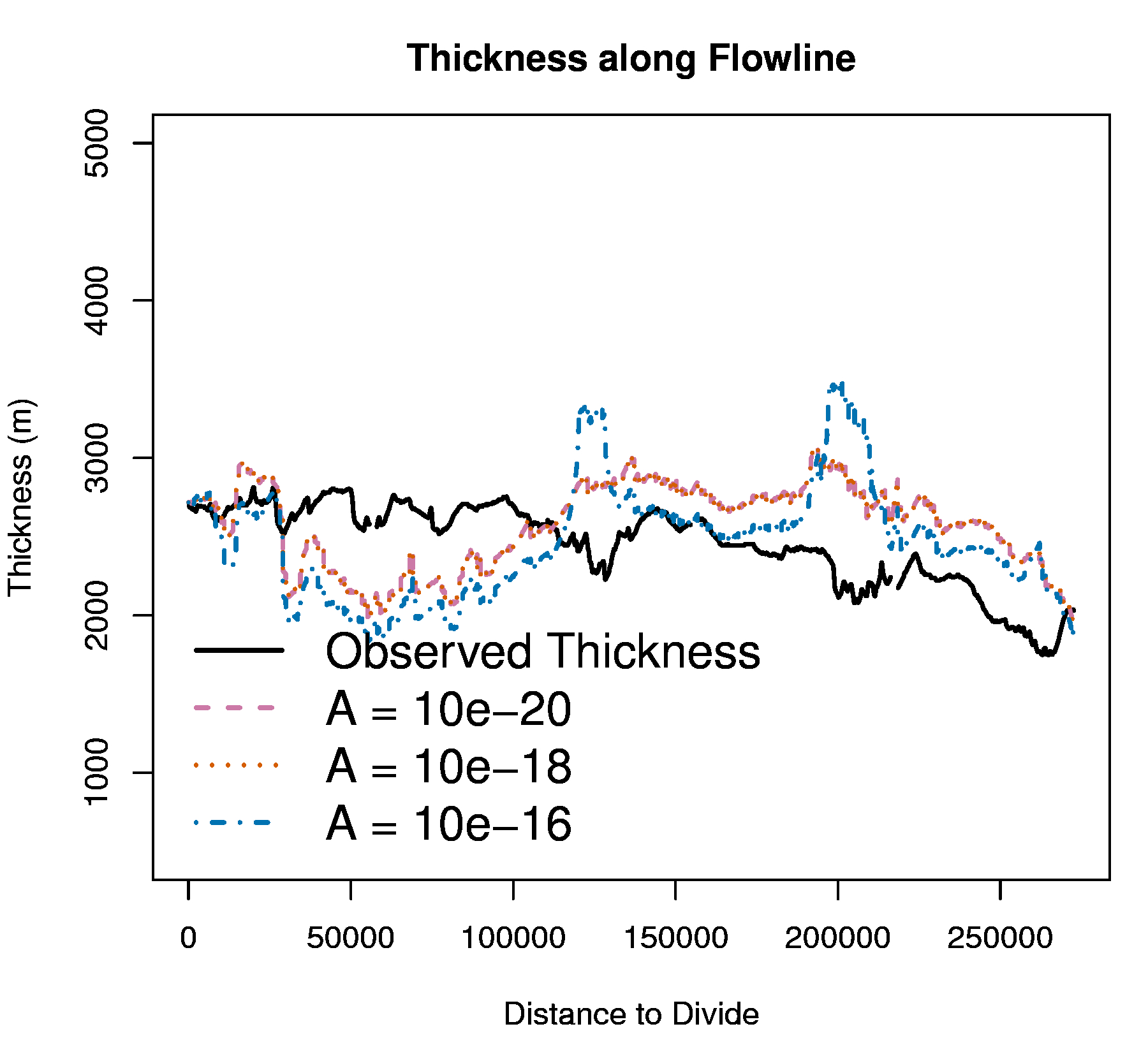}}
		\vspace{0mm}
		\subfloat[Solutions corresponding to the narrowest flowline width (red line in Figure \ref{fig:flowlineb})]{\includegraphics[width=0.5\textwidth,clip, trim=0cm 0cm 0cm 1.2cm]{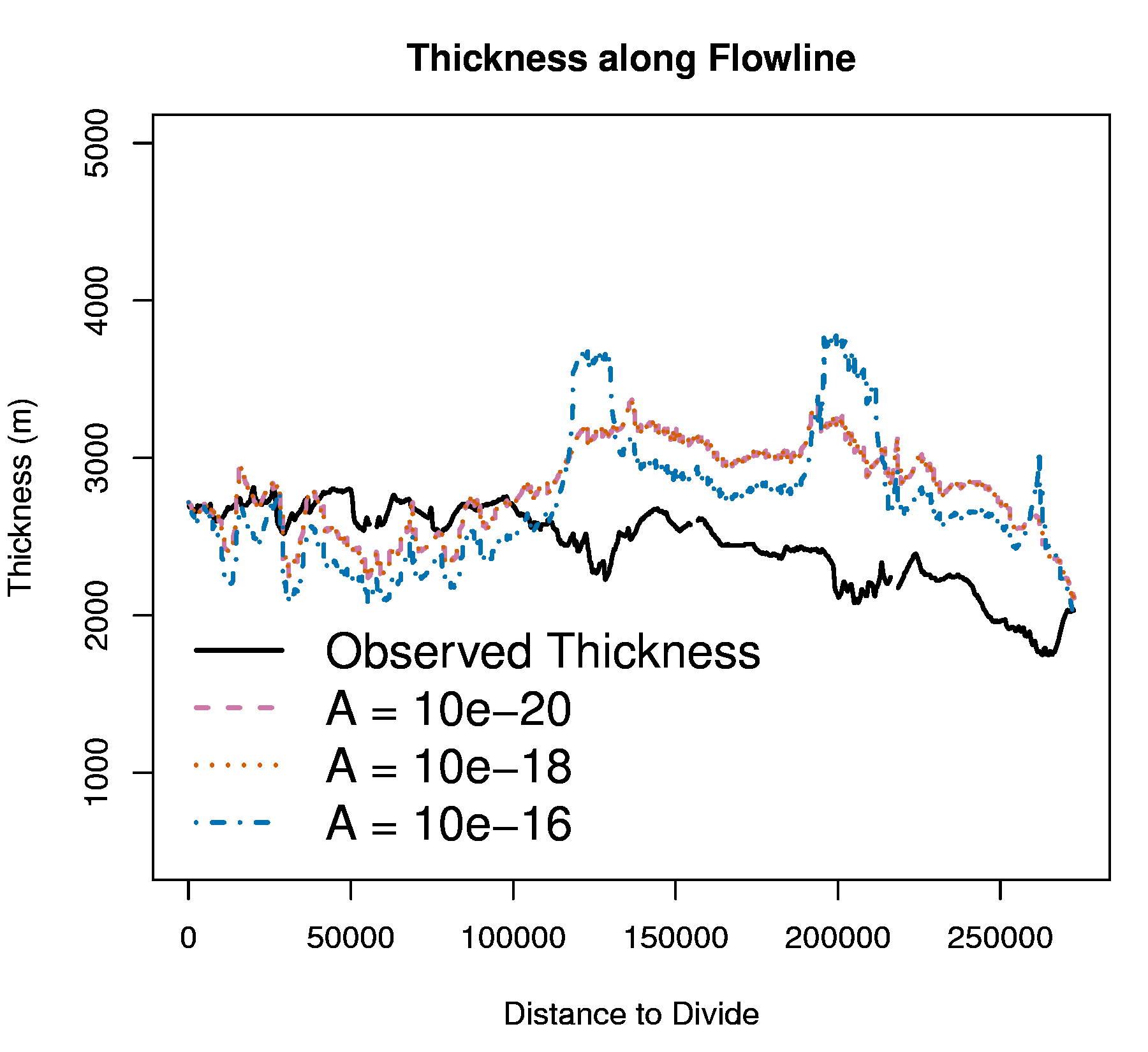}}
		\caption{Thwaites Glacier thickness solutions corresponding to different flowline widths and rheologic coefficient values. The non-statistical approach provides deterministic ice thickness solutions without uncertainty estimates, and none of these solutions seem to reconstruct the ice thickness well.}
		\label{fig:nonstat}
	\end{figure}

	We first describe a naive non-statistical approach for solving ice thickness using (\ref{eqn:sia}) and (\ref{eqn:velcorrection}). This involves finite differences and solving fifth order polynomials. Typically, $ a(x) $ and $ \tau(x) $ are taken to be well observed surface quantities, as are the surface velocity and surface elevation. For a fixed value of $ A $ and a plug-in estimate of $\omega$, we can obtain the flux $ \bar{v}(x)h(x)\omega(x) $ on a predetermined grid along the $ x$-axis by solving the following ordinary differential equation derived from (\ref{eqn:sia}),
	\begin{eqnarray}\label{eqn:thick}
	\begin{aligned}
	\bar{v}(x)h(x)\omega(x) - \bar{v}(x_0)h(x_0)\omega(x_0) &= \int_{x_0}^{x} (a(s)-{\tau}(s))\omega(s)ds,\\
	\text{and \hspace{4mm}} \bar{v}(x_0)h(x_0)\omega(x_0) &= C_0 \text{\hspace{4mm} is a given initial value.}
	\end{aligned}
	\end{eqnarray}
	Plugging in \eqref{eqn:velcorrection} into \eqref{eqn:thick}, we obtain a fifth order polynomial of $h(x)$. Using a numerical solver, we can compute the ice thickness as a function of other variables; denote this as $M\left(v_s(x), s(x), a, \tau, \omega, A, h_0\right)$. Here, we write $h_0 = h(x_0)$ as the initial thickness value for simplicity.
	However, the solutions for the system may not be unique or do not exist for some locations. For locations where solutions exist, we select the ones that are closest to the observed ice thickness. The results from this non-statistical inversion method using plugged-in estimates of the flow width from Figure \ref{fig:flowlineb} are shown in Figure \ref{fig:nonstat}; we see that none of these solutions seem to reconstruct the ice thickness well, for instance, some sections along the flowline have large departures, up to 1 km, from the observed thickness. Also the bedrock topography is not recognized by the reconstructed ice thickness, for example, the thin ice near 130 km, 200 km and 250 km cannot be reproduced from the naive method. Therefore, we conclude that although this non-statistical approach provides fast and rough estimates for the unknown, there are a few drawbacks that make it undesirable. First, the plug-in estimates for $\omega$ are chosen informally; this choice is not based on a well-defined criterion. Figure \ref{fig:flowlinea} shows a few possible streamlines estimated by tracing the direction of surface velocity vector field using the MATLAB toolbox Ice Flowlines \citep{flowline}. For every pair of streamlines at each side of the centerline, we can take the transverse distance as the flow width. However, depending on different starting upstream locations, the widths can differ substantially. Second, measurement errors in observations have not been accounted for, which may lead to over-confident estimates. Third, without quantifying uncertainty associated with the estimates, any conclusions obtained would be unreliable.

	These drawbacks motivate us to propose a statistical approach that can integrate multiple sources of information, multiple data sets, underlying physical law and error sources, in a single framework to provide parameter estimates along with its uncertainty and probabilistic prediction for unobserved ice thickness.

	\subsection{Data}\label{sec:data}
	Several surface data sets are available for the WAIS. These include (1) surface elevation measured by using Airborne Topographic Mapper (ATM) instrumentation as part of Operation IceBridge funded aircraft survey campaigns \citep{surfe}, (2) surface velocity with spatial resolution of 450 m as part of the NASA Making Earth System Data Records for Use in Research Environments (MEaSUREs) Program \citep{RignotVel} and (3) estimates of ice accumulation rate with spatial resolution $ \sim $ 55 km from \cite{van2006reassessment}. We first smooth the surface elevation, then derive the surface slope $ s(x) $ by taking central differences. Surface elevation has been smoothed, because the glacier dynamics model is very sensitive to small variations in the surface slope. Based on the physical understanding of the glacier, the small variations in surface slope are not realistic and mostly due to observational errors. For the Thwaites Glacier, we also have ice thickness observations from airborne radar with $ \sim $ 14 m spacing along the flowline of total length 272.8 km \citep{ThickDat}.  Note that surface velocity, surface slope, ice accumulation and thinning rates have different spatial resolutions; velocity and slope are interpolated to the locations where we have thickness data, and accumulation and thinning rate are interpolated to the same grid as the flowline width (details described in Section \ref{sec:inputprocessmodel}). Our methods would adapt in straightforward fashion to situations where the data are on different resolutions; the only additional burden is computational. All the data that can be observed or derived from observable quantities are summarized in Table \ref{tab:data}.
	
	\section{Hierarchical approach}\label{sec:3}
	In this section, we introduce a hierarchical approach that can integrate the multiple data sets summarized in Table \ref{tab:data} and the glacier dynamics model described by (\ref{eqn:sia}) and (\ref{eqn:velcorrection}) to infer ice thickness along the flowline. This framework has the flexibility of incorporating multiple error sources, including the discrepancy between the physical model and the true underlying process, multiple sources of observational errors and error due to discretization of the glacier dynamics model. We discuss the computational challenges of inference based on the full hierarchical approach. Computational considerations require us to approximate this approach. Our main goal is to (1) estimate the rheologic coefficient $ A $ and quantify its uncertainty due to various sources of data and the physical model, and (2) predict the ice thickness incorporating multiple sources of information.

	\subsection{Model overview}
\begin{figure}[ht]
		\centering
		\includegraphics[width=0.9\textwidth]{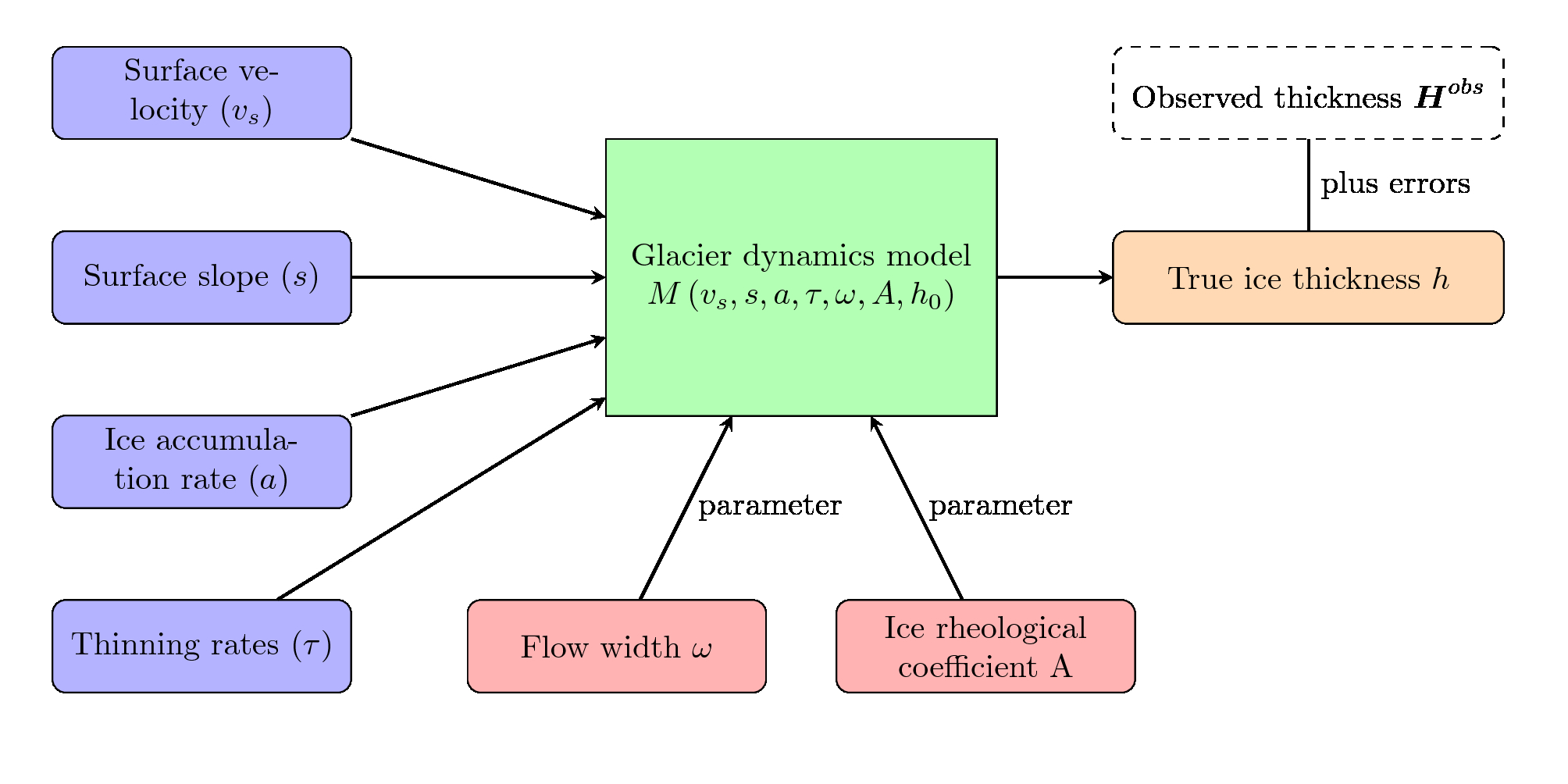}
		\caption{Dynamics model $M$ is based on Equations (\ref{eqn:sia}) and (\ref{eqn:velcorrection}). It links input processes, model parameters and ice thickness.}
		\label{fig:flowchart}
	\end{figure}

	Given the need to combine multiple sources of data with a physical model, a hierarchical Bayesian approach is natural. Here we introduce the notation and begin with an overview of the statistical model. We then describe each component in details in the following subsections. Let $ \Gamma \subseteq R$ denote the continuous spatial domain along the flowline with total length of 272.8 km from the divide to upstream of the grounding line. At a location $ x\in\Gamma $, we have the true processes $ v_s(x)$ for surface velocity,  $ s(x) $ for surface slope, $ a(x) $ for ice accumulation rate, $\tau(x)$ for thinning rate, and $ h(x) $ for ice thickness. These processes obey the physical relationship described by the glacier dynamics model (see Figure \ref{fig:flowchart}), in which rheologic coefficient $ A $, and flowline width process $\omega$, are unknowns to be estimated. Let $\bs{x} = \left( x_1,\dots,x_n \right)^T \in\Gamma$ denote a vector of locations where we have ice thickness observations, $\bs{H}^\text{obs} = \left( H_{x_1}^\text{obs},\dots,H_{x_n}^\text{obs} \right)^T  $ be a vector of observed thickness, and $ \bs{h} = \left( h({x_1}),\dots,h({x_n})\right)^T$ be the true process evaluated at $\bs{x}$. Similarly, observed inputs are denoted as $\bs{V}_s^{\text{obs}},\bs{S}^{\text{obs}},\bs{a}^{\text{obs}}$, and $\bs{\tau}^{\text{obs}} $. Note that the input observations may be from different spatial locations.
	
	We formulate the statistical model in the following hierarchical form.\begin{align}
	&\text{\textbf{Ice thickness Model: }}  \nonumber\\
	\label{eqn:datamodel}
	&\text{\hspace{6mm} observations model: } \bs{H}^{\text{obs}}\mid \bs{h}, \bs{\theta} \sim
	N(\bs{h},\sigma^2_H\bs{I}), \\
	\label{eqn:processmodel}
	&\text{\hspace{6mm} physics (deterministic) model: }  h \mid v_s,s,a,\tau,\omega, \bs{\theta} = M\left(v_s,s,a,\tau,\omega, A, h_0 \right).\\\label{eqn:flowwidthmodel}
	&\text{\textbf{Flowline Width Model: }} \omega\mid\bs{\theta} \sim \text{GP} \left(C(\bs{\theta}_\omega) \right).\\
	\nonumber
	&\text{\textbf{Input Process Model: }}  v_s,s,a,\tau\mid \bs{\theta}, \bs{V}_s^{\text{obs}},\bs{S}^{\text{obs}},\bs{a}^{\text{obs}},\bs{\tau}^{\text{obs}}\sim \\
	\nonumber & \hspace{45mm}f(v_s\mid\bs{V}_s^{\text{obs}},\bs{\theta}_v)f(s\mid\bs{S}^{\text{obs}},\bs{\theta}_s)f(a\mid\bs{a}^{\text{obs}},\bs{\theta}_{a})f(\tau\mid\bs{\tau}^{\text{obs}},\bs{\theta}_{\tau}),\\
	\nonumber
	&\text{\hspace{6mm} velocity model: } f(v_s\mid\bs{V}_s^{\text{obs}},\bs{\theta}_v),  \\
	\nonumber
	&\text{\hspace{6mm} slope model: }  f(s\mid\bs{S}^{\text{obs}},\bs{\theta}_s),  \\
	\nonumber
	&\text{\hspace{6mm} accumulation rate model: }  f(a\mid\bs{a}^{\text{obs}},\bs{\theta}_{a}),  \\
	\nonumber
	&\text{\hspace{6mm} thinning rate model: }  f(\tau\mid\bs{\tau}^{\text{obs}},\bs{\theta}_{\tau}).  \\\nonumber
	&\text{\textbf{Prior: }}  p(\bs{\theta}) = p(\sigma^2_H)p(h_0)p( A)p( \bs{\theta}_\omega)p(\bs{\theta}_v)p(\bs{\theta}_s) p(\bs{\theta}_{a})p(\bs{\theta}_{\tau}),\end{align}
	where $ \bs{\theta} = ( \sigma^2_H, h_0, A, \bs{\theta}_\omega, \bs{\theta}_v, \bs{\theta}_s, \bs{\theta}_{a},\bs{\theta}_{\tau})^T $ includes all the unknown parameters in the model. The ice thickness observations model (\ref{eqn:datamodel}) implies conditional independence for the observations, with errors determined by $ \sigma^2_H $. The physics model (\ref{eqn:processmodel}) imposes physical constraints for multiple processes: conditional on the initial thickness value $ h_0 $, rheologic coefficient $  A $, the flow width $ \omega $ and the input processes (surface velocity, surface slope, ice accumulation and thinning rates), the thickness process is related to the input processes through the deterministic glacier dynamics model $M$. The flow width process $ \omega $ is modeled with a Gaussian process (\ref{eqn:flowwidthmodel}) with a Mat\'en covariance function $ C(\bs{\theta}_\omega ) $ where $ \bs{\theta}_\omega = (\nu, \sigma^2_\omega,\phi,\tau^2)^T $; we fix the smoothness parameter $ \nu=1.5 $ to reflect that the width is smooth based on expert opinion. The input processes, however, are unknown and assumed to be independent, conditional on their respective model parameters $ \bs{\theta}_v, \bs{\theta}_s, \bs{\theta}_{a}, \bs{\theta}_{\tau} $ and observations. We now provide details for the components of the hierarchical model and comment on their extensions.

	\subsection{Ice thickness model}\label{sec:modelfordata}
	The $ n $-dimensional thickness observations $ \bs{H}^{\text{obs}} $, conditional on the underlying true thickness process $ \bs{h} $, are modeled as independent normal variables centered at the true values with unknown variance $ \sigma^2_H $. A simple error model is justified since its main source is the instrumental error in acquiring the thickness data. The true thickness at the observation locations have a deterministic relationship with the true surface velocity, surface slope, accumulation rate and thinning rate as described in the glacier dynamics model (\ref{eqn:sia}) and (\ref{eqn:velcorrection}), which balances the change in flux $\left(\frac{\partial}{\partial x} \left( v_s(x)-\frac{A}{20}\left(\rho g|s\left(x\right)|\right)^3h(x)^4 \right) h(x)\omega(x)\right)$ with the net change in depth $\left((a(x)-\tau{(x)})\omega(x)\right)$ for mass conservation.

	Conditional on the initial thickness value $h_0$, the rheologic coefficient $ A $, the flowline width $ \omega(x)$ and the input processes ($ v_s(x), s(x), a(x), \text{and } \tau(x) $), solving for thickness $h(x)$  involves two steps: (i) integrating $(a(x)-\tau{(x)})\omega(x)$ from the divide (initial location used when solving the differential equation) $x_0$ to an observation location $x_j\in\bs{x}$, and (ii) computing the flux, $\bar{v}(x_j)h(x_j)\omega(x_j)$, at $x_j$ and solving for ice thickness.

	Here we describe the details for steps (i) and (ii) listed immediately above. Step (i) integration over the spatial domain can be approximated with a finite sum since, in practice, we only obtain the continuous processes flow width, accumulation and thinning rates at finite locations. Let $ \bs{x}^{g} = ({x}^{g}_0,\dots,{x}^{g}_m)^T \in\Gamma $ be a collection of locations; here, we take $ \bs{x}^{g} $ on a grid with $ {x}^{g}_0 = {x}_0 $ and $ {x}^{g}_m > {x}_n  $ to cover the study domain. We approximate the integral $ \int_{x_0}^{x_j} (a(s)-\tau(s))\omega(s)ds $ by a sum of the products evaluated at $ \bs{x}^{g}$, $ \sum_{i=0}^{I}\left( a(x^{g}_{i,i+1/2})-\tau(x^{g}_{i,i+1/2})\right) \omega(x^{g}_{i,i+1/2})\triangle_{{i},{i+1}} $, from $ x^{g}_0 $ to $ x^{g}_I $, where the interval $[x^{g}_I, x^{g}_{I+1}] $ contains the observation location  $x_j$, $ a(x^{g}_{i,i+1/2}) $ is the average of $a(x^{g}_{i})$ and $ a(x^{g}_{i+1}) $, $ \tau(x^{g}_{i,i+1/2}) $ is the average of $\tau(x^{g}_{i})$ and $ \tau(x^{g}_{i+1}) $, $ \omega(x^{g}_{i,i+1/2}) $ is the average of $\omega(x^{g}_{i})$ and $ \omega(x^{g}_{i+1}) $, and $ \triangle_{{i},{i+1}} $ is the distance between locations $ x^{g}_{i}$ and $x^{g}_{i+1} $. Note that the grid $ \bs{x}^{g} $ can be different from the observation locations $ \bs{x} $, and there is a trade off between accuracy and computational efficiency in determining its dimension: high (low) resolution $ \bs{x}^{g} $ increase (decrease) approximation accuracy and reduce (increase) efficiency. An error term can be added to the process model to capture the discretization error, for instance, $ \int_{x_0}^{x_j} \left( a(s)-\tau(s)\right) \omega(s)ds = \sum_{i=0}^{I}\left( a(x^{g}_{i,i+1/2})-\tau(x^{g}_{i,i+1/2})\right) \omega(x^{g}_{i,i+1/2})\triangle_{{i},{i+1}} + \epsilon(x) $. However, we neglect $ \epsilon(x) $ here, because accumulation rate, thinning rate and width are smooth processes based on physical understanding; in this case, the discretization error is small. Moreover, the Gaussian process model for flow width is very flexible; it may adjust itself to reduce model discrepancy. We repeat step (i) for all observation locations $ x_j, j =1,\dots,n $. In step (ii) we compute the flux $\bar{v}(x_j)h(x_j)\omega(x_j) \approx \bar{v}(x_0)h(x_0)\omega(x_0)+ \sum_{i=0}^{I}\left( a(x^{g}_{i,i+1/2})-\tau(x^{g}_{i,i+1/2})\right) \omega(x^{g}_{i,i+1/2})\triangle_{{i},{i+1}}$ for each observation location $x_j$. Denote the flux quantities as $F_j, j = 1,\dots,n$. Then, we solve the below fifth order polynomial for ice thickness,
	\begin{equation}\label{eqn:poly}
	F_j -\left( v_s(x_j) - \frac{A}{20} \left( \rho g s(x_j)\right) ^3  h(x_j)^4 \right) h(x_j)\omega(x_j)=0.
	\end{equation}

	\subsection{Flowline width model}\label{sec:widthprior}
	The flowline width $ \omega(x), x\in\Gamma $ is not observed, however, we know it is relatively smooth based on the estimates by tracing the direction of velocity vector field and our physical understanding of glaciers \citep{docquier2014thwaites}. We impose smoothness to flowline width by modeling it with a Gaussian process. Here, we choose a Mat\'ern covariance function with smoothness $\nu = 1.5$, since it produces a relatively smooth process. Therefore, the flowline width $ \bs{\omega} = \left( \omega(x_0^g),\dots,\omega(x_m^g)\right) ^T $ at the grid $ \bs{x}^g $ has a multivariate normal distribution with covariance matrix $ \Sigma$. The $ (i,j) $ element of the covariance matrix, $\Sigma_{ij} $, is the covariance between $ \omega(x_i^g) $ and $ \omega(x_j^g)  $, and $ \Sigma_{ij}  = C(\mid\mid x_i^g-x_j^g\mid\mid) = \sigma_\omega^2\left(1 +\frac{\sqrt{3}\mid\mid x_i^g-x_j^g\mid\mid}{\phi} \right) \exp(-\frac{\sqrt{3}\mid\mid x_i^g-x_j^g\mid\mid}{\phi}) $ for $ i\ne j $, or $\Sigma_{ij} = \sigma_\omega^2 + \tau^2 $ for $ i=j $ \citep{RasmussenGPM}. Other models for flow width that preserve the smoothness of the process and can interpolate in between observations, for instance, B-spline basis function and Gaussian process with a smoother covariance function,  are also reasonable for this application.

	Based on the observed flowline width, we know it is wider upstream and narrower downstream;  this knowledge is incorporated in the flowline width model by specifying a mean function for the Gaussian process. In the glacier dynamics model (\ref{eqn:sia}), we can assume a value for the rheologic coefficient $ A = 0 $ and plug in the narrowest observed flowline width (red dashed line in Figure \ref{fig:flowlineb}) to evaluate the integral from the divide to the thickness observation locations $ \bs{x} $ assuming there are no observational errors, we then solve for the glacier width at $ \bs{x} $ using the observed ice thickness $ \bs{H}^{obs}  $ and (\ref{eqn:poly}). A linear interpolation of the glacier width at $ \bs{x} $ is then the mean function for the Gaussian process. Based on our simulated examples, the choice of A value and the observed flowline width to deduce the mean function does not affect the results of the Bayesian model.

	\subsection{Input process model}\label{sec:inputprocessmodel}
	The multiple data sets are implicitly integrated via their respective true processes through the glacier dynamics model in (\ref{eqn:sia}). We model the input processes (surface velocity, surface slope, ice accumulation and thinning rates) as independent processes conditional on their respective parameters ($ \bs{\theta}_v, \bs{\theta}_s, \bs{\theta}_{a}, \bs{\theta}_{\tau}$). Our assumption does not imply that individual input processes are marginally independent, rather we assume the small-scale variation in the true processes, conditional on the smoothed observations, are independent. Let $ \bs{V}_s^\text{obs}, \bs{S}^\text{obs}$, $\bs{a}^\text{obs}$ and $\bs{\tau}^\text{obs}$ be vectors of observed surface velocity, surface slope, ice accumulation and thinning rates, respectively. Each individual process is related to their respective observations by the following.
	\begin{align*}
	\text{velocity model: }& v_s(x)\mid \bs{\theta}_v , \bs{V}_s^{\text{obs}} 
	= S_{\lambda_v} \bs{V}_s^{\text{obs}}  + \epsilon_v(x),
	\hspace{4mm}\epsilon_v(x) \sim N(0,\sigma_v^2),\\
	\text{slope model: }& s(x)\mid \bs{\theta}_s , \bs{S}^{\text{obs}} 
	= S_{\lambda_s}\bs{S}^\text{obs} + \epsilon_s(x),
	\hspace{4mm}\epsilon_s(x) \sim N(0,\sigma_v^2),\\
	\text{accumulation rate model: }& a(x)\mid \bs{\theta}_v , \bs{a}^{\text{obs}} 
	= S_{\lambda_{a}}\bs{a}^{\text{obs}} + \epsilon_{a}(x),
	\hspace{4mm}\epsilon_{a}(x) \sim N(0,\sigma_{a}^2),\\
	\text{thinning rate model: }& \tau(x)\mid \bs{\theta}_v , \bs{\tau}^{\text{obs}}  
	= S_{\lambda_{\tau}}\bs{\tau}^{\text{obs}} + \epsilon_{\tau}(x),
	\hspace{4mm}\epsilon_{\tau}(x) \sim N(0,\sigma_{\tau}^2),
	\end{align*}
	where $ S_{\lambda_v},  S_{\lambda_s} $, $ S_{\lambda_{a}} $ and $ S_{\lambda_{\tau}} $ are smoothing matrices, for instance smoothing splines \citep{nychka1988smoothingsplines}, and $ \epsilon_v(x), \epsilon_s(x) $, $ \epsilon_{a}(x)$ and $ \epsilon_{\tau}(x)$ are small-scale variation in the true processes not captured by the smoothed observations. In practice, surface observations are obtained from various sources and do not share the same locations; therefore, they are often interpolated to the same locations before entering physical models. The input process models allow for observations with different spatial resolutions.

	In the glacier dynamics model, we evaluate the accumulation and thinning rates at location $ \bs{x}^g $, $ \bs{a} = (a(x^g_0),\dots,a(x^g_m))^T $ and $ \bs{\tau} = (\tau(x^g_0),\dots,\tau(x^g_m))^T $, to approximate the integral using step (i) in  Section \ref{sec:modelfordata}, and we evaluate surface velocity and slope at observation locations $ \bs{x} $, denoted by $ \bs{v}_s = (v_s(x_1),\dots,v_s(x_n))^T $ and $ \bs{s} = (s(x_1),\dots,s(x_n))^T  $, respectively, to solve for thickness using step (ii) in Section \ref{sec:modelfordata}. Therefore, the dimension of the input process models $ \bs{v}_s,\bs{s},\bs{a}, \bs{\tau}\mid\bs{\theta}$ is about $ 2(n+m) $.

	\subsection{Model inference and computational details}\label{sec: model inference}
	Maximum likelihood inference for the hierarchical model is difficult. The challenge is that the glacier dynamics model can not be solved analytically; its solutions can only be obtained through numerical approximation. The likelihood function is hence not in closed form; therefore, its maximization requires sophisticated optimization over a high-dimensional parameter space.  Bayesian inference based on the posterior distribution $ \pi(\bs{\theta}, \bs{v}_s,\bs{s},\bs{a},\bs{\tau}, \bs{\omega}\mid \bs{H}^{obs},\bs{V}_s^{obs},\bs{S}^{obs},\bs{a}^{obs}, \bs{\tau}^{obs}) $ is not difficult in principle.

	However, the computational cost of inference for the hierarchical model poses a major challenge in practice. To handle the computational challenge, two compromises are made in the hierarchical model: (1) we pre-smooth the input processes using smoothing splines and treat them as the true processes for solving the ice thickness in the glacier dynamics model, and (2) instead of estimating the range parameter $ \phi $ for the flowline width, we estimate it from the observed flowline width. After these two modifications to the full hierarchical model, its posterior distribution reduces to $ \pi(A, \sigma^2_{\omega}, h_0, \tau^2, \bs{\omega}\mid \bs{H}^{obs}, \bs{V}_s^{obs}, \bs{S}^{obs}, \bs{a}^{obs}, \bs{\tau}^{obs})$. We use Markov Chain Monte Carlo (MCMC) with Metropolis-Hasting updates to sample from the posterior distribution. We use a uniform prior with a range from $ 0 $ to $ 10^{-16} $ for the rheologic coefficient $ A $; the range is selected based on expert opinion. We use inverse gamma priors for variance $ \sigma^2_\omega $  (IVG(2,$ 10^8 $)) and $ \tau^2 $  (IVG(2,$ 10^4 $)), based on exploratory analysis as the follow: for fixed values of rheologic coefficient, if we assume that the errors in thickness observations are negligible, then we can derive flow width using (\ref{eqn:thick}) and (\ref{eqn:poly}); we explore the variation in the flowline width by solving it using several values of $ A $ and thickness observations with added errors.

	Fixing the range parameter at a reasonable value reduces computational expenses and we find in practice that this approach does not affect the results significantly; also the main purpose of assuming a spatial process here is to interpolate assuming some smoothness. 
	Based on arbitrarily chosen flow width as shown in Figure \ref{fig:flowlineb}, we estimated $\hat{\phi}=40,000$ (m). In our implementation, we solve the high-order polynomial (\ref{eqn:poly}) using the R package rootSolve \citep{rootsolve}. Each MCMC is run from three dispersed starting values; based on the trace plots from several MCMC runs, we decide that 200,000 MCMC iterations suffice.
	\section{Thwaites glacier}\label{sec:4}
	 We first apply the hierarchical approach described in Section \ref{sec:3} to synthetic observations simulated from the physical model. This allows us to investigate the following:  (1) inference and prediction performance, (2) the skill of recovering the local features of thickness by combining the physical model and surface observations, and (3) the number of thickness observations needed to provide useful predictions. Then, we apply the model to the Thwaites glacier with 5, 10 and 25 thickness observations, about 46 km, 30 km and 14 km apart in distance, respectively along the flowline.

	\subsection{Simulated example}
\begin{figure}[!htbp]
		\centering
		\includegraphics[width=0.7\textwidth,clip, trim=0cm 0cm 0cm 1.2cm]{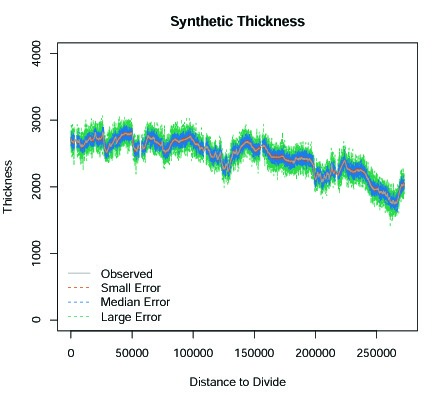}
		\caption{Synthetic ice thickness data with different magnitude of observational errors. Small, medium and large errors have standard deviations of 10, 50 and 100 m, respectively.}
		\label{fig:synthick}
	\end{figure}

	To preserve the topographic features of the ice thickness, we treat the observed thickness on Thwaites glacier as the true underlying process and simulate surface velocity from the physical model (\ref{eqn:sia}). We then fit the hierarchical model using a small number of synthetic thickness observations generated from the synthetic true thickness with different magnitude of \textit{i.i.d} normal error (Figure \ref{fig:synthick}). This model fitting procedures are repeated for 5, 10, and 25 training locations. We then compare the performance in estimating the rheologic coefficient $ A $, and predicting the ice thickness. To make synthetic data match realistic observations, we also use the observed surface slope, ice accumulation rate, thinning rate and the observed flow width corresponding to the red dashed line in Figure \ref{fig:flowlineb} for data generation. We let the true rheologic coefficient to be $ A = 10^{-18} $.

\begin{figure}[!htbp]
		\centering
		\subfloat[]{\includegraphics[width=0.35\textwidth,clip, trim=0cm 0cm 0cm 1.2cm]{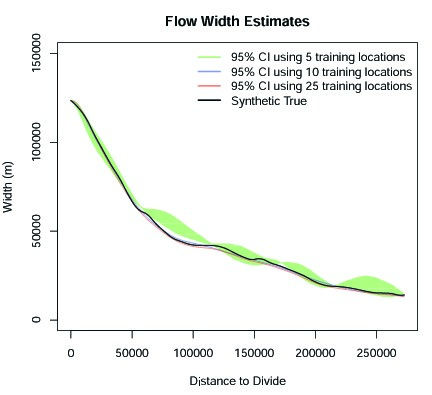}}
		\subfloat[]{\includegraphics[width=0.35\textwidth,clip, trim=0cm 0cm 0cm 1.2cm]{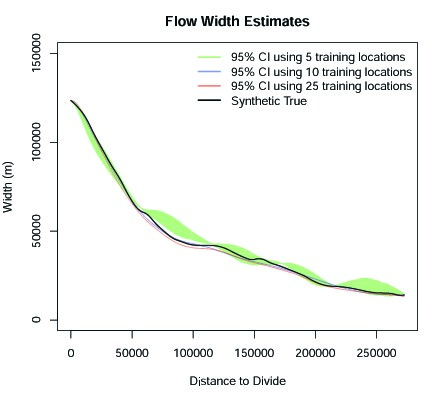}}
		\subfloat[]{\includegraphics[width=0.35\textwidth,clip, trim=0cm 0cm 0cm 1.2cm]{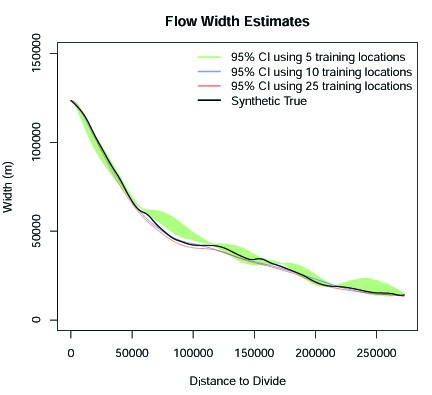}}
		\caption{Predicted flow widths using 5, 10 and 25 synthetic ice thickness observations. Flow width predictions using small (a), medium (b) and large (c) magnitude of observational errors. The magnitude of observational errors does not change the shape of the width. Larger number of training locations produces tighter flow width that is more comparable to the synthetic true width.}
		\label{fig:Synwidth}
	\end{figure}

\begin{figure}[!htbp]
		\centering
		\captionsetup[subfigure]{width=0.3\textwidth}
		\subfloat[Use 5 synthetic observations with small error]{\includegraphics[width=0.35\textwidth,clip, trim=0cm 0cm 0cm 6cm]{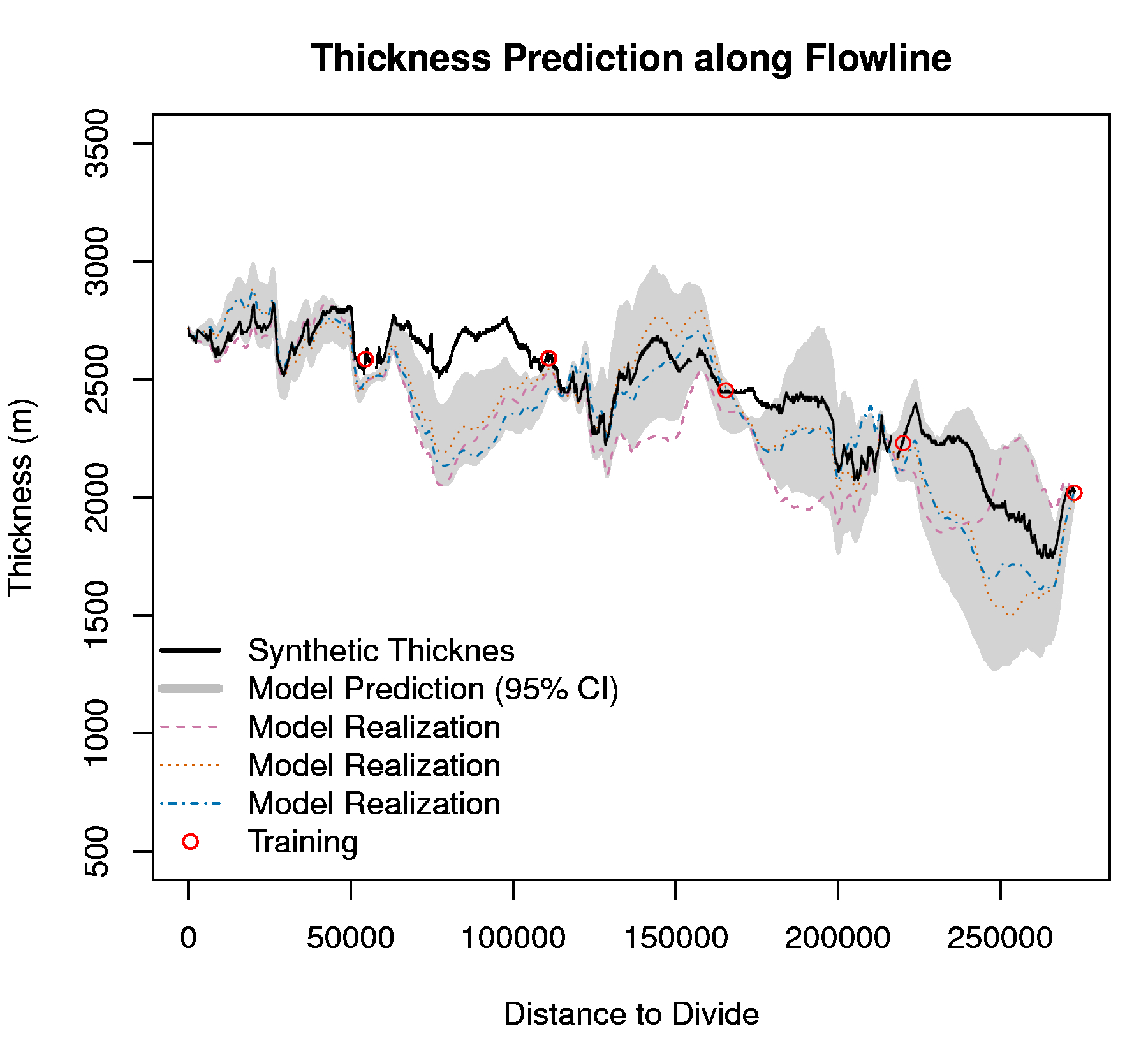}}
		\subfloat[Use 5 synthetic observations with medium error]{\includegraphics[width=0.35\textwidth,clip, trim=0cm 0cm 0cm 6cm]{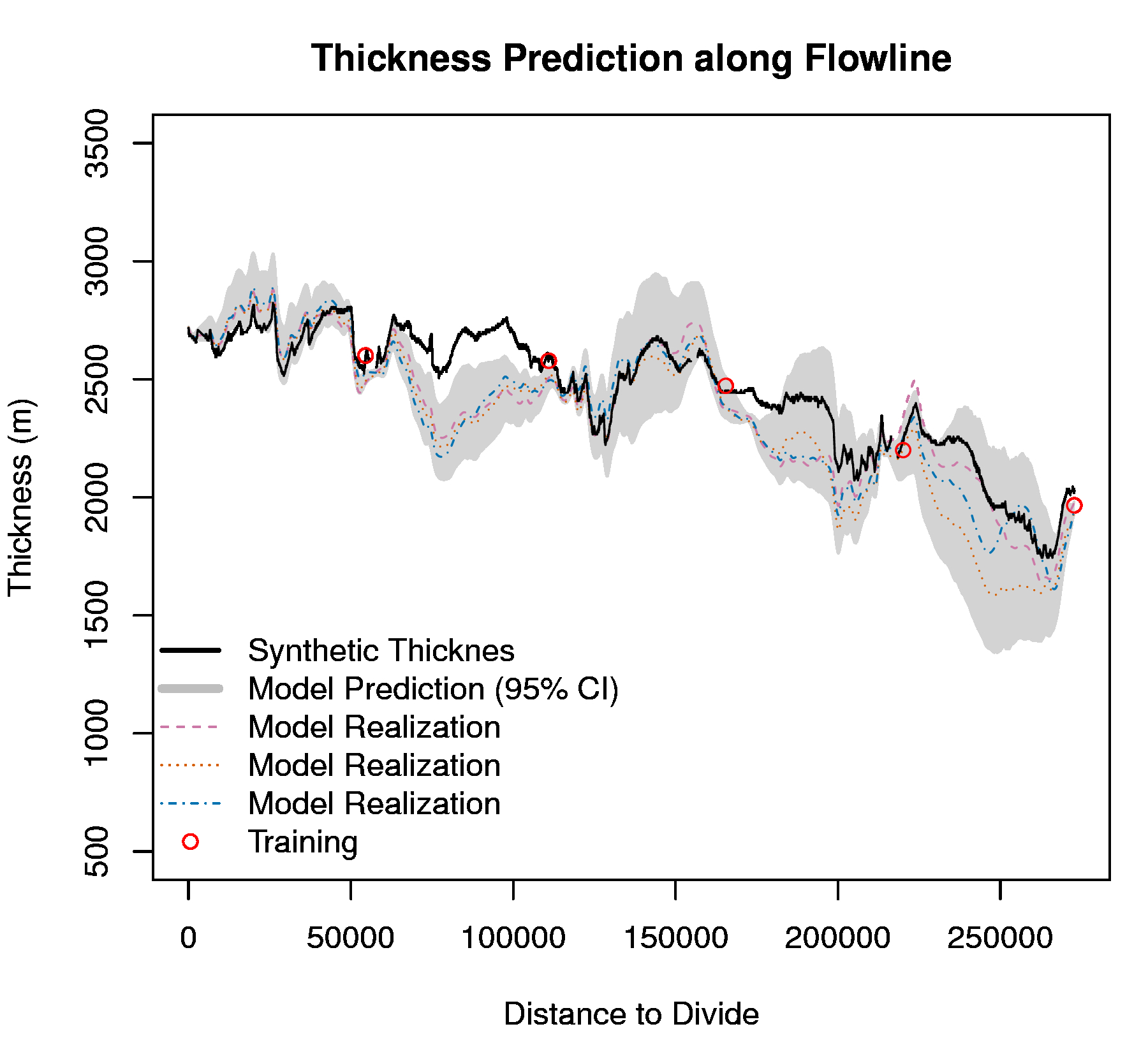}}
		\subfloat[Use 5 synthetic observations with large error]{\includegraphics[width=0.35\textwidth,clip, trim=0cm 0cm 0cm 6cm]{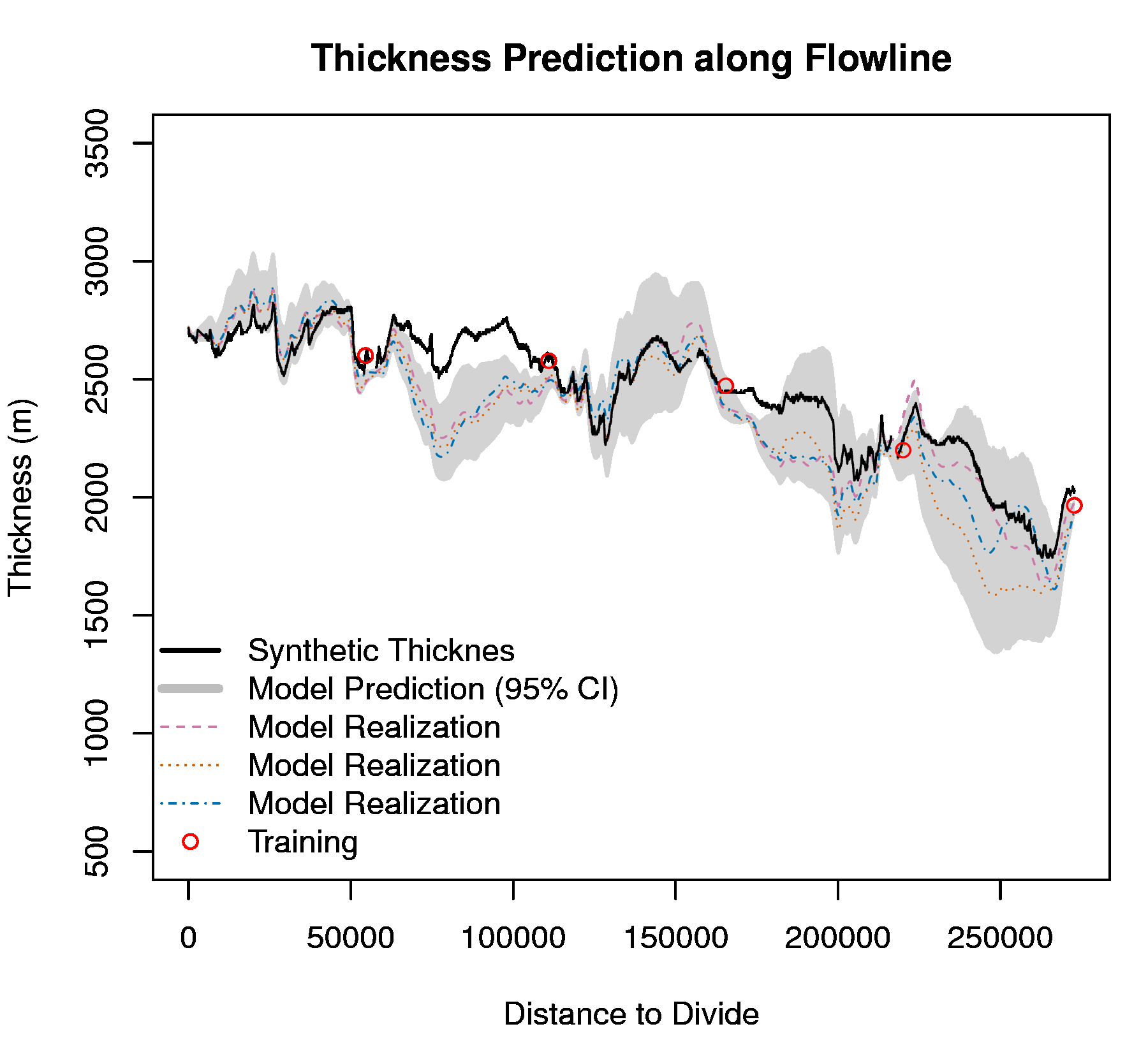}}
		\vspace{0mm}
		\subfloat[Use 10 synthetic observations with small error]{\includegraphics[width=0.35\textwidth,clip, trim=0cm 0cm 0cm 6cm]{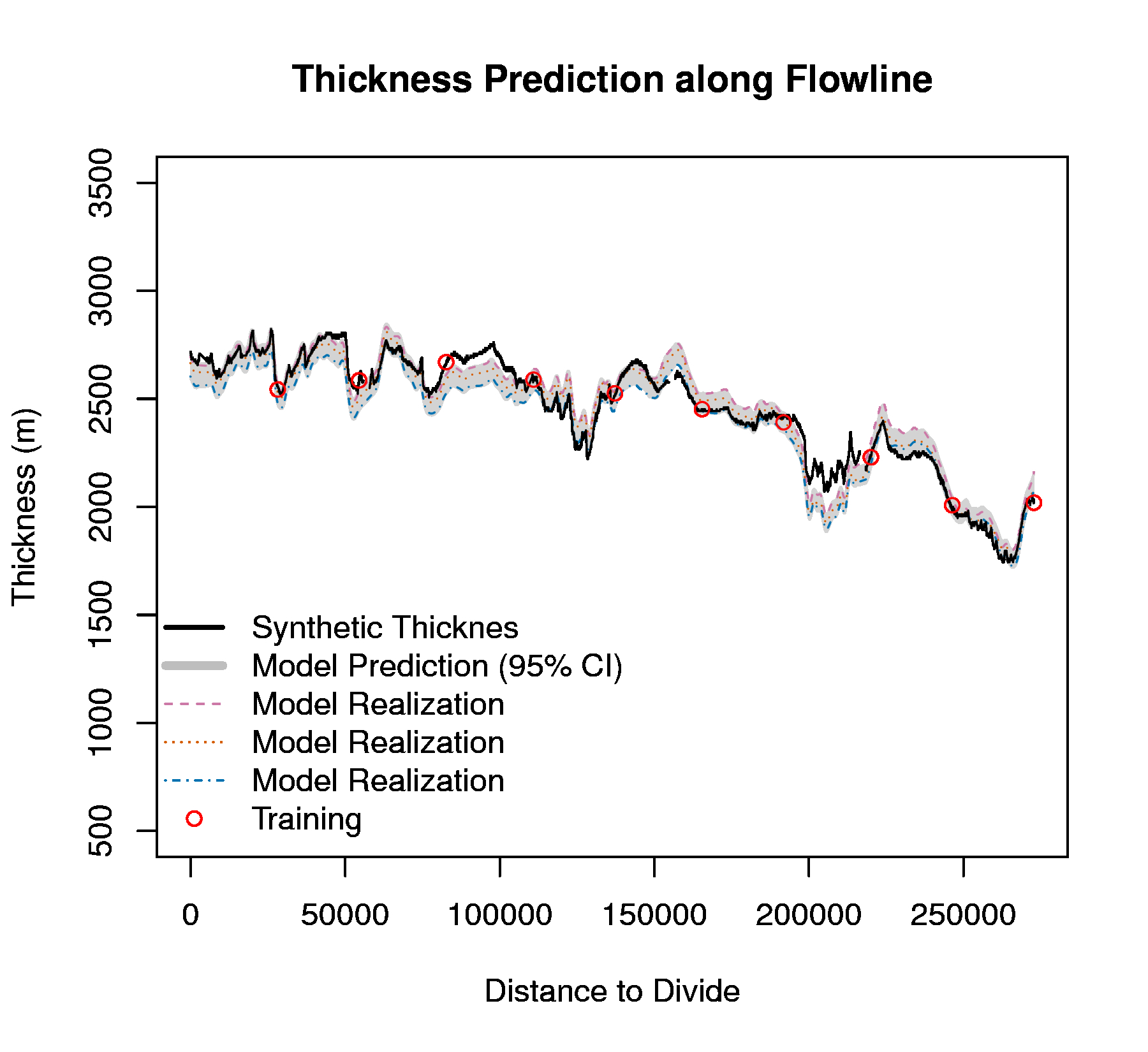}}
		\subfloat[Use 10 synthetic observations with medium error]{\includegraphics[width=0.35\textwidth,clip, trim=0cm 0cm 0cm 6cm]{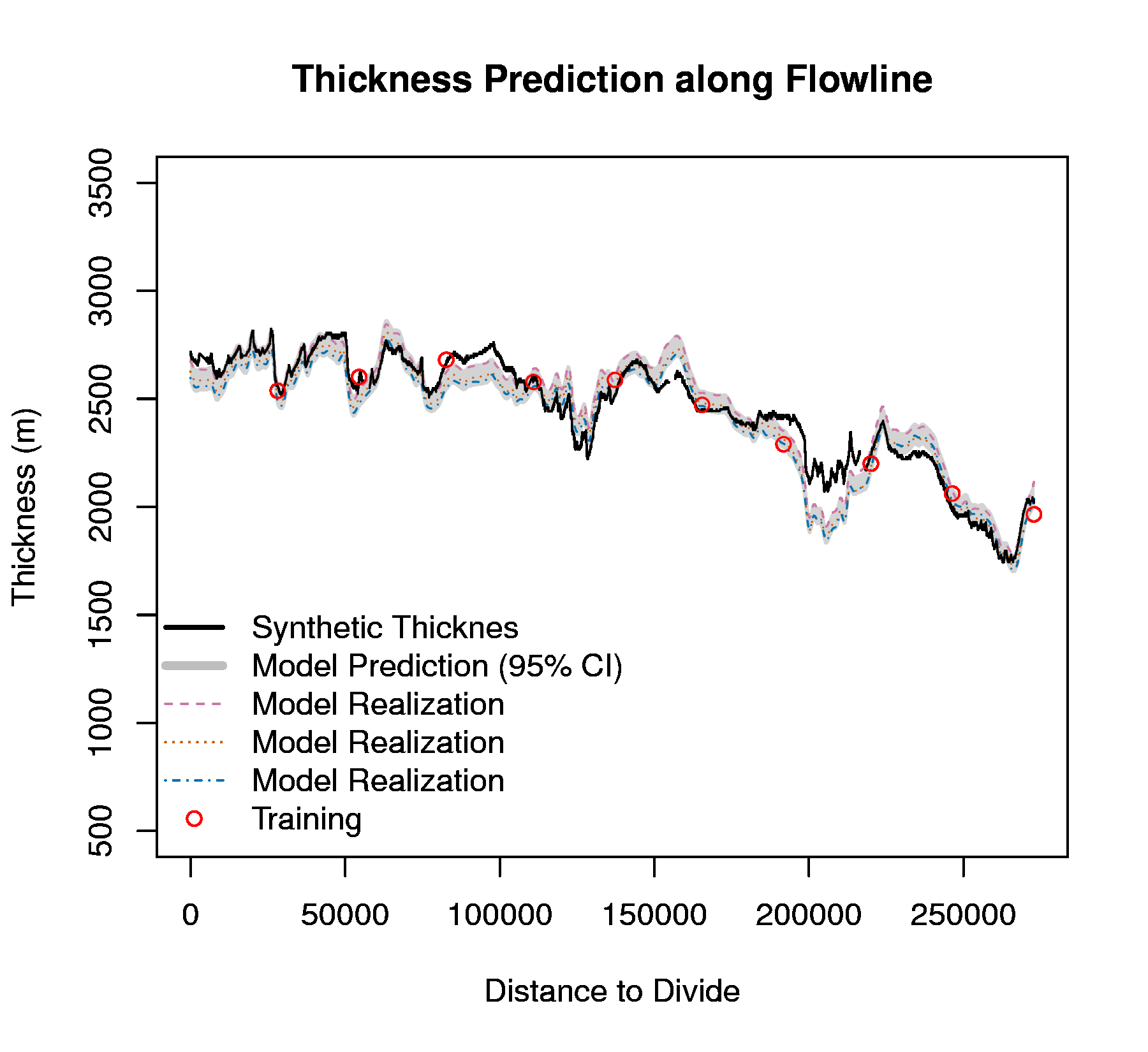}}
		\subfloat[Use 10 synthetic observations with large error]{\includegraphics[width=0.35\textwidth,clip, trim=0cm 0cm 0cm 6cm]{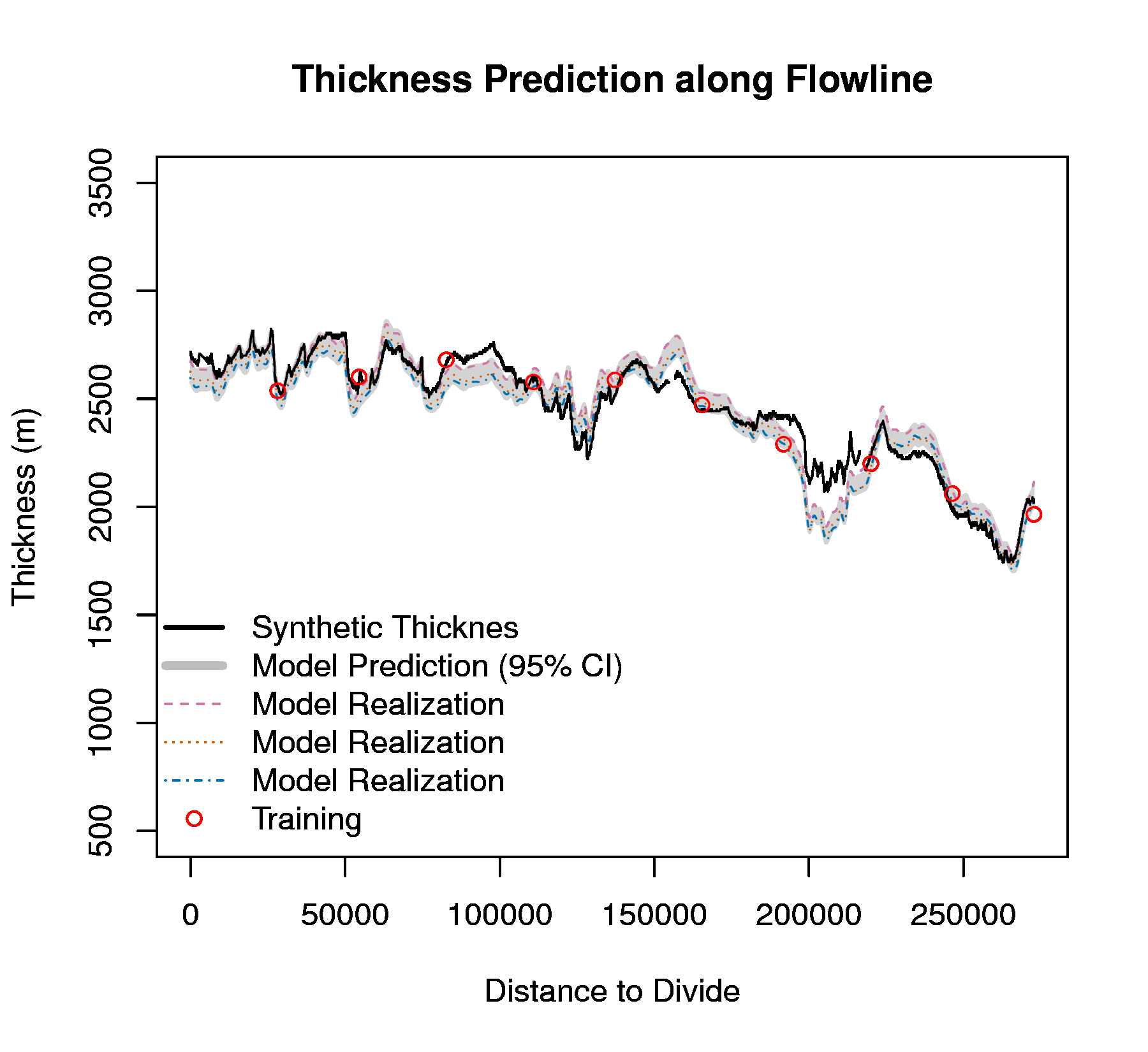}}
		\vspace{0mm}
		\subfloat[Use 25 synthetic observations with small error]{\includegraphics[width=0.35\textwidth,clip, trim=0cm 0cm 0cm 6cm]{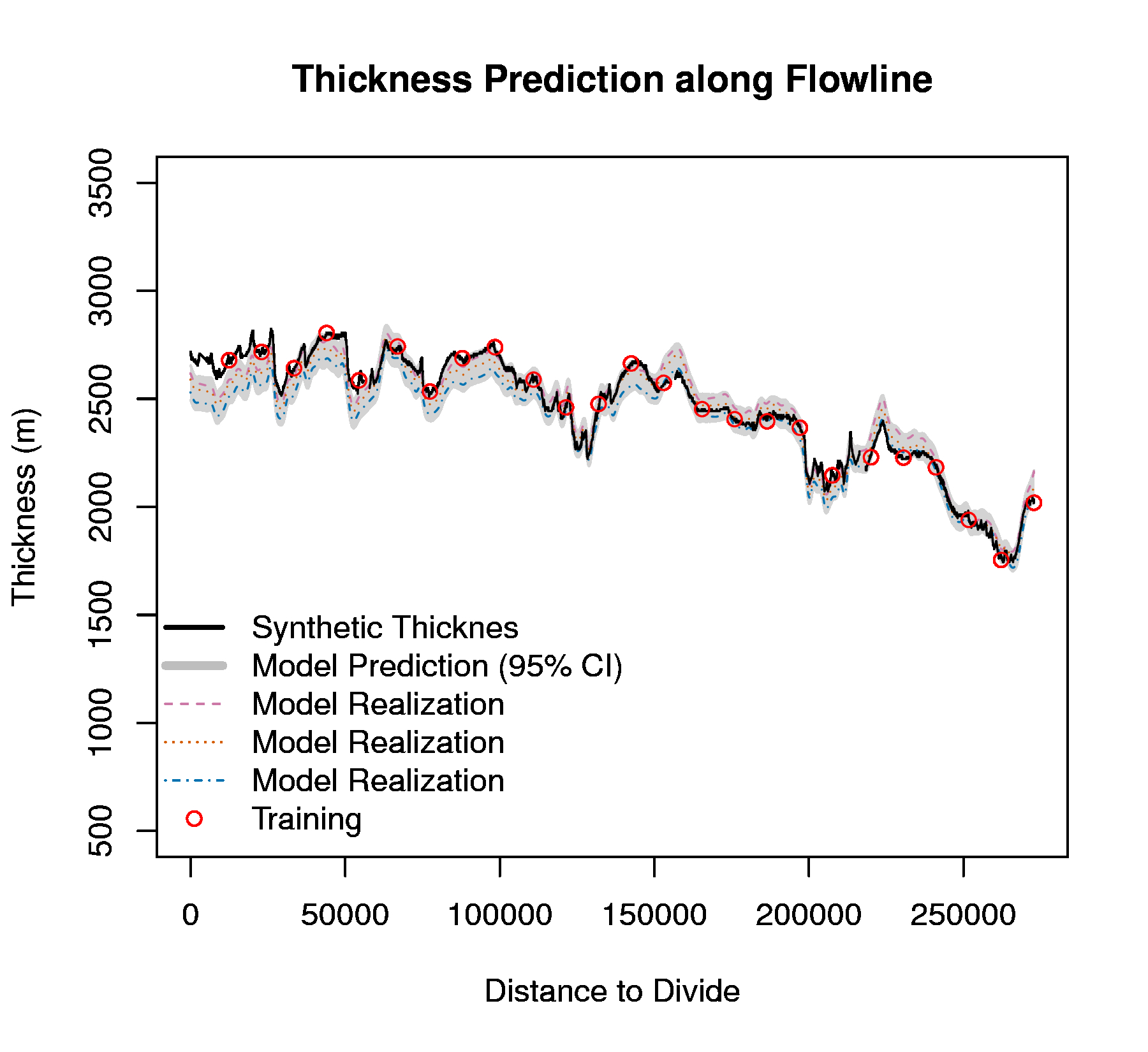}}
		\subfloat[Use 25 synthetic observations with medium error]{\includegraphics[width=0.35\textwidth,clip, trim=0cm 0cm 0cm 6cm]{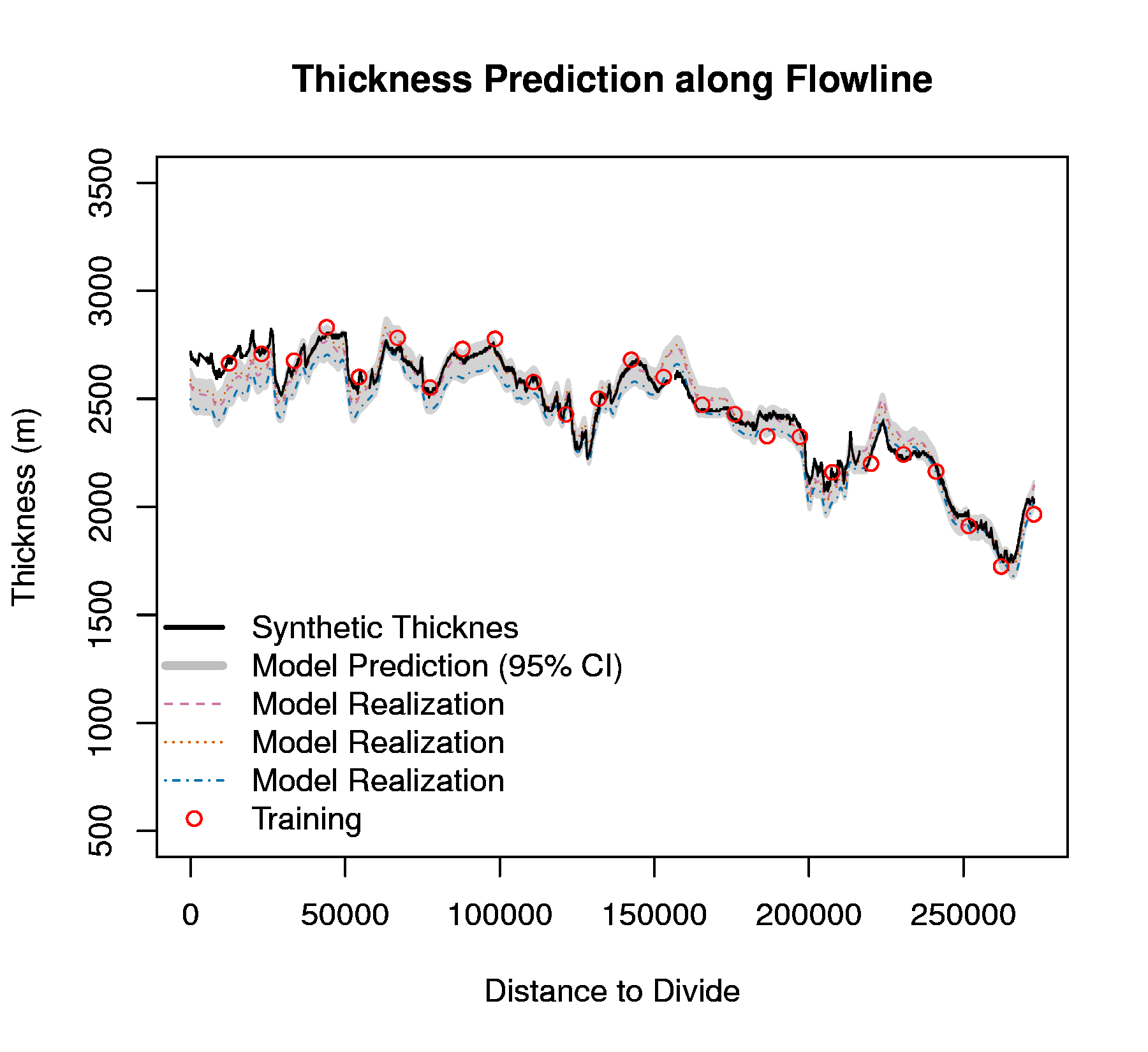}}
		\subfloat[Use 25 synthetic observations with large error]{\includegraphics[width=0.35\textwidth,clip, trim=0cm 0cm 0cm 6cm]{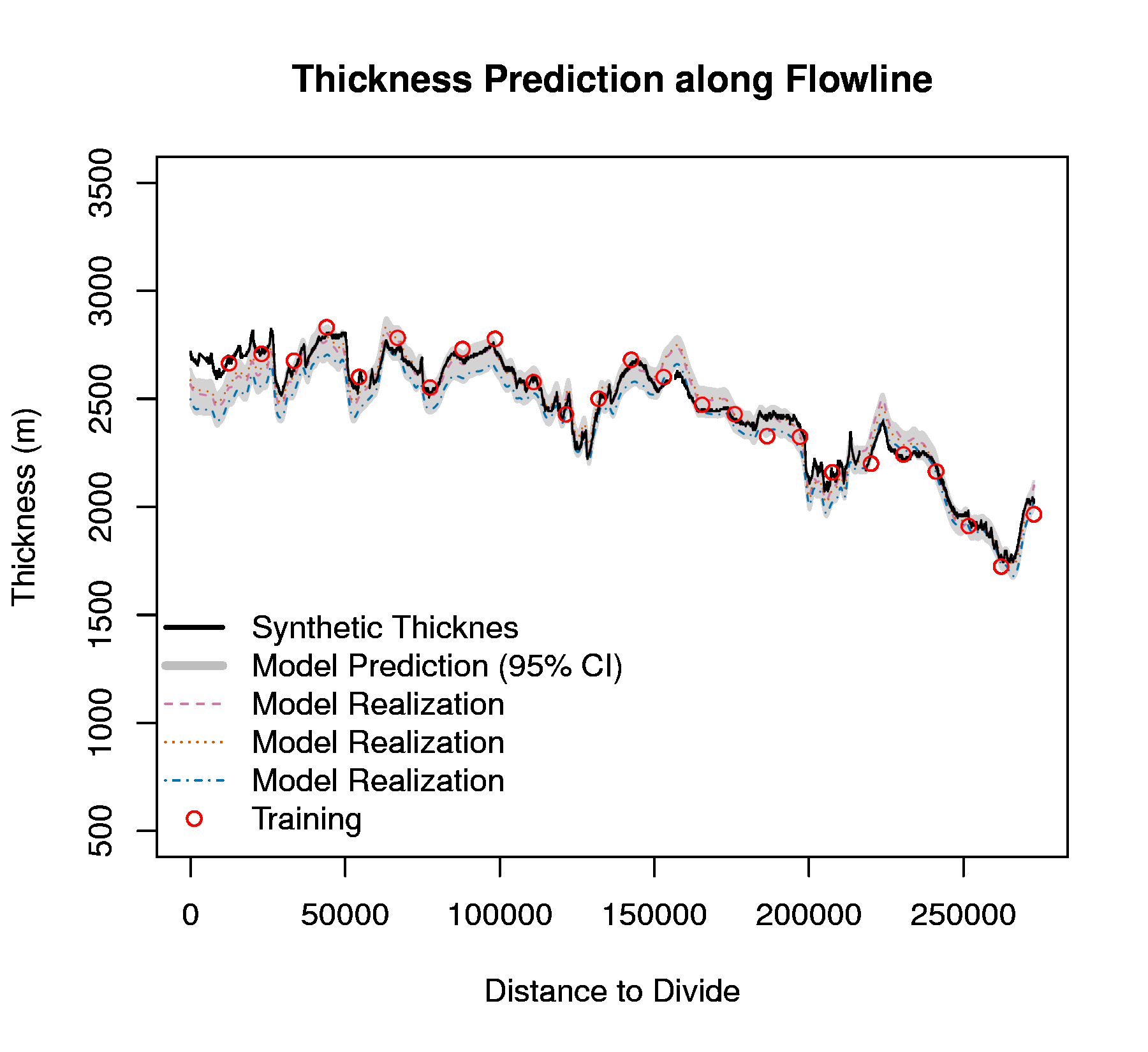}}
		\caption{Predicted ice thickness from flow model using 5, 10, and 25 synthetic thickness observations. Prediction results are similar under different error schemes, and a larger number of training locations provide more certain thickness prediction.}
		\label{fig:Synthick}
	\end{figure}

	The inference on the rheologic coefficient $ A $ under the three error schemes are similar: posterior means are all around $ 5.07\times 10^{-17} $, and 95\% credible intervals decrease for increasing training locations, for instance, 95\% CI's are [$ 2.085\times10^{-18}, 9.683\times10^{-17} $] and [$ 3.002\times10^{-18}, 9.632\times10^{-17}$] for 5 and 25 training locations (under the medium error scheme), respectively. The predicted flow width for the entire flowline is shown in Figure \ref{fig:Synwidth}. We see that as the number of observations increases, the uncertainties associated with the width predictions decrease dramatically. We also study the 95\% credible intervals from the width predictions; the coverages, shown in Table \ref{tab:cov}, should be around 95\%, however, are substantially lower than the nominal as the number of training locations increases. The thickness predictions from the physical model are shown in Figure \ref{fig:Synthick}. From the few model realizations, we see that the fine scale structure is preserved from the prediction, even when using only 5 training locations. As the number of training locations increases, the predicted thickness is tighter around the synthetic true thickness. However, the predicted thickness using 5 training locations has a larger uncertainty as its location gets further from the divide. 
	The increasing uncertainty in thickness prediction is due to the ODE solution for the flux in (\ref{eqn:thick}); in approximating the integral from the divide to downstream, small departures from the width estimate to the synthetic true are accumulated along the distance. This reflects the sensitivity of ODE solution to errors, and we could reduce the uncertainty by solving the system using smaller segments along the glacier. The uncertainty also reduces as the number of training locations increases, results from better width estimation. It seems that estimating the observational error in thickness using our approach is very difficult; observational error $ \sigma_H $ has posterior means around 5,000 m for all three error schemes. This is mostly likely a result from the discretization in approximating the integral and simplification of observational error model. A practical implication of our results above is that thickness observational errors does not have substantial impact in thickness reconstruction, because much of the information about ice thickness comes from the physical model and well observed surface observations.
	
	\subsection{Application}
\begin{figure}[!htbp]
		\centering
		\subfloat[]{\includegraphics[width=0.35\textwidth,clip, trim=0cm 0cm 0cm 6cm]{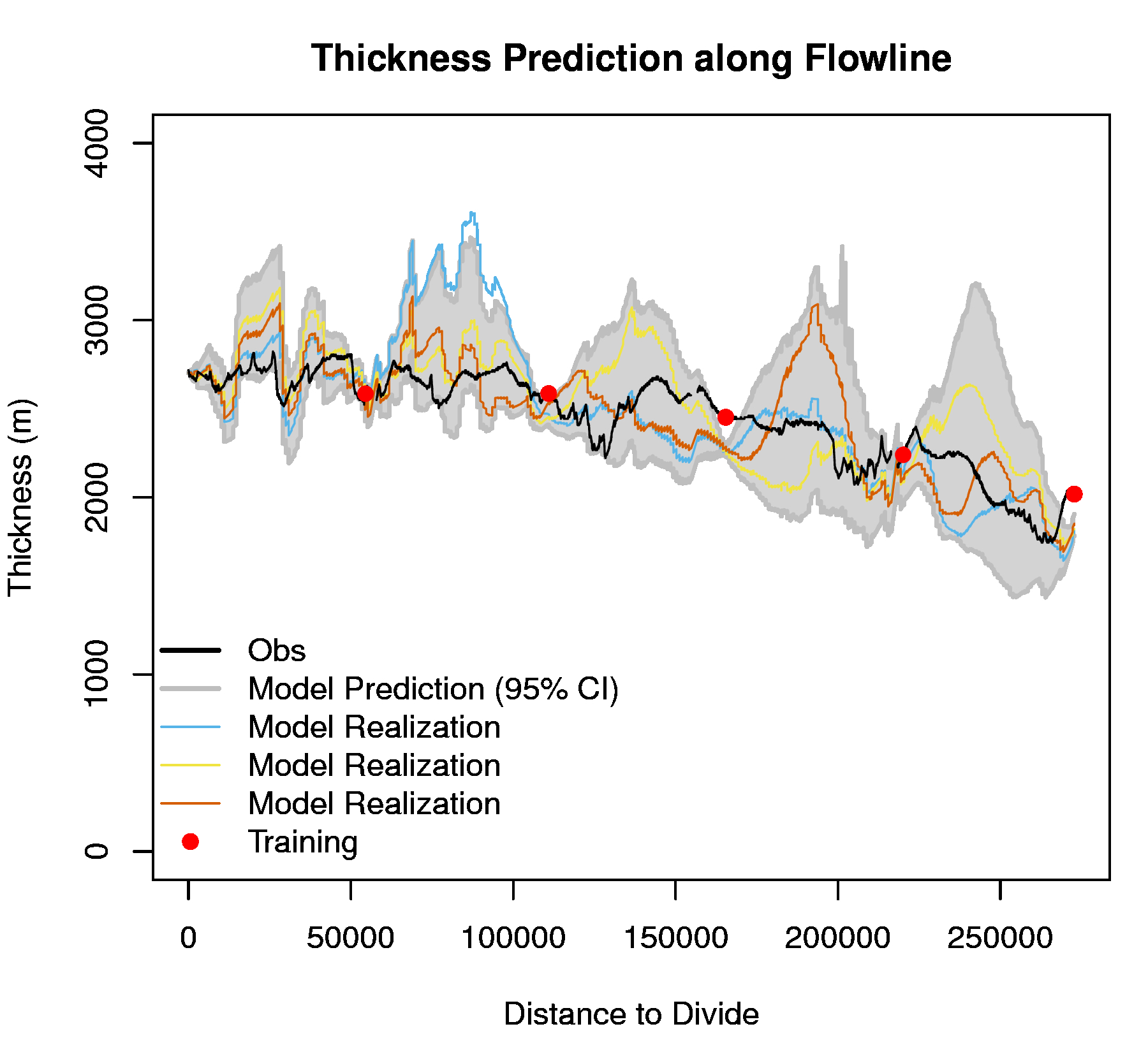}}
		\subfloat[]{\includegraphics[width=0.35\textwidth,clip, trim=0cm 0cm 0cm 6cm]{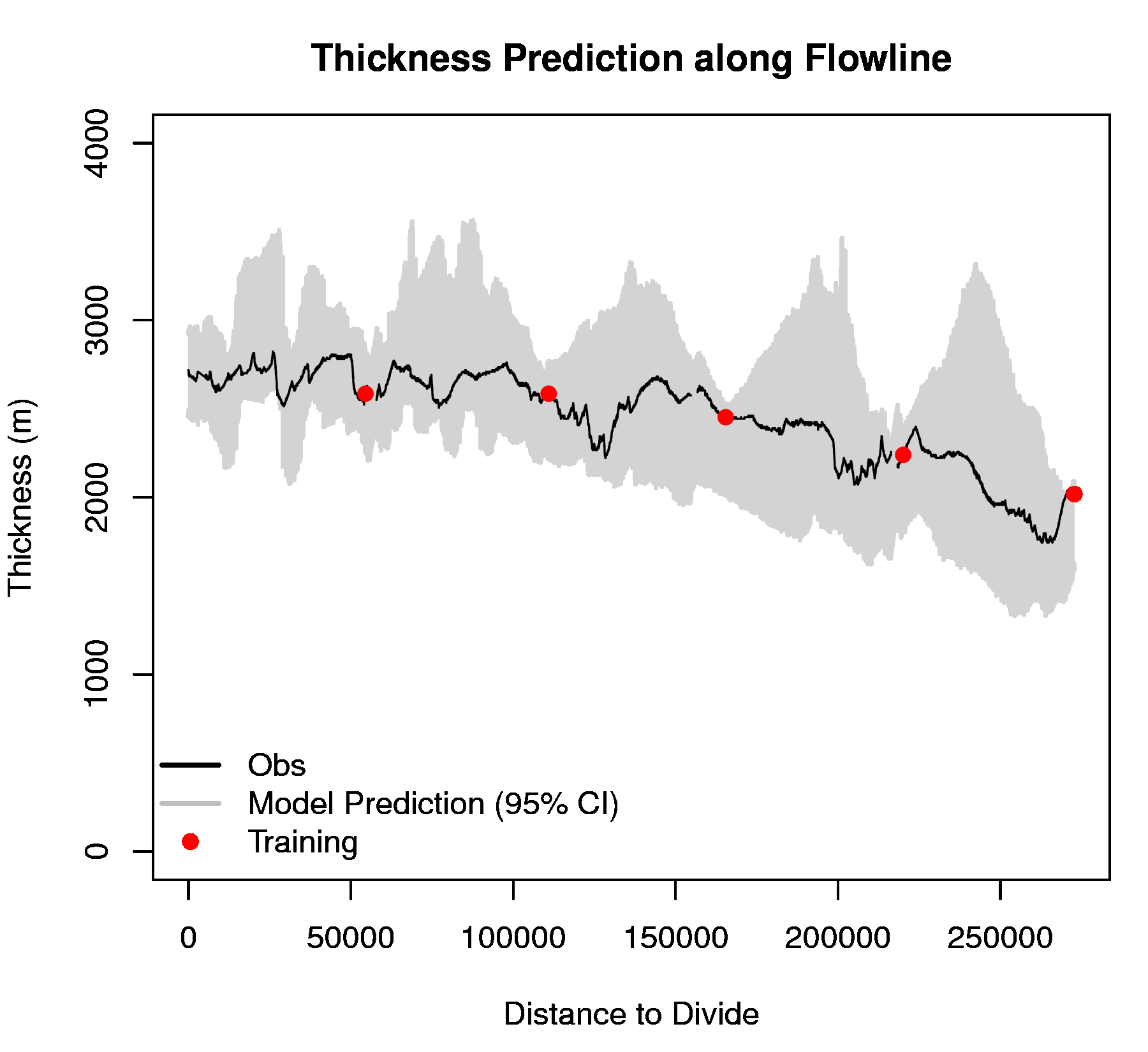}}
		\vspace{0mm}
		\subfloat[]{\includegraphics[width=0.35\textwidth,clip, trim=0cm 0cm 0cm 6cm]{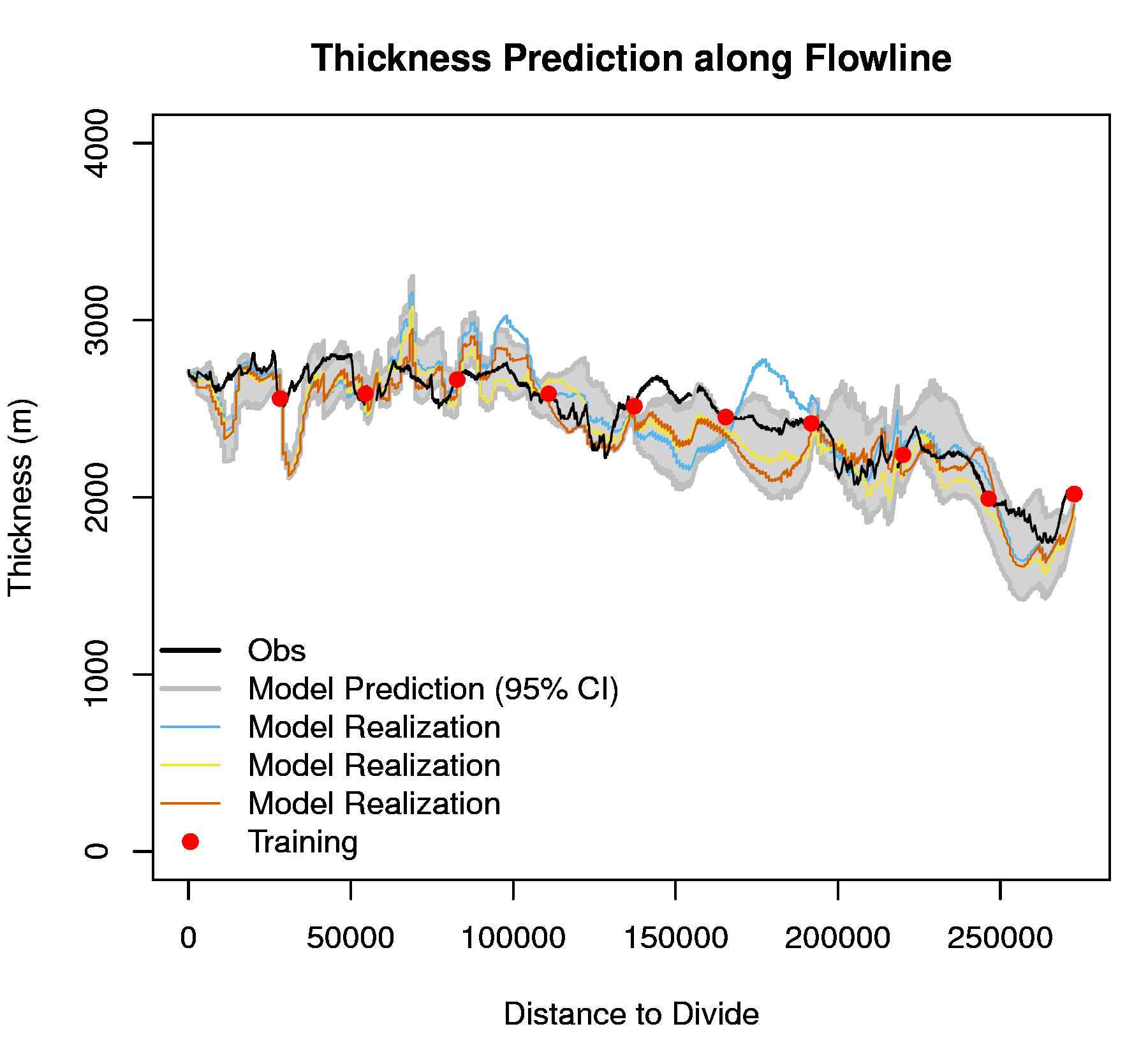}}
		\subfloat[]{\includegraphics[width=0.35\textwidth,clip, trim=0cm 0cm 0cm 6cm]{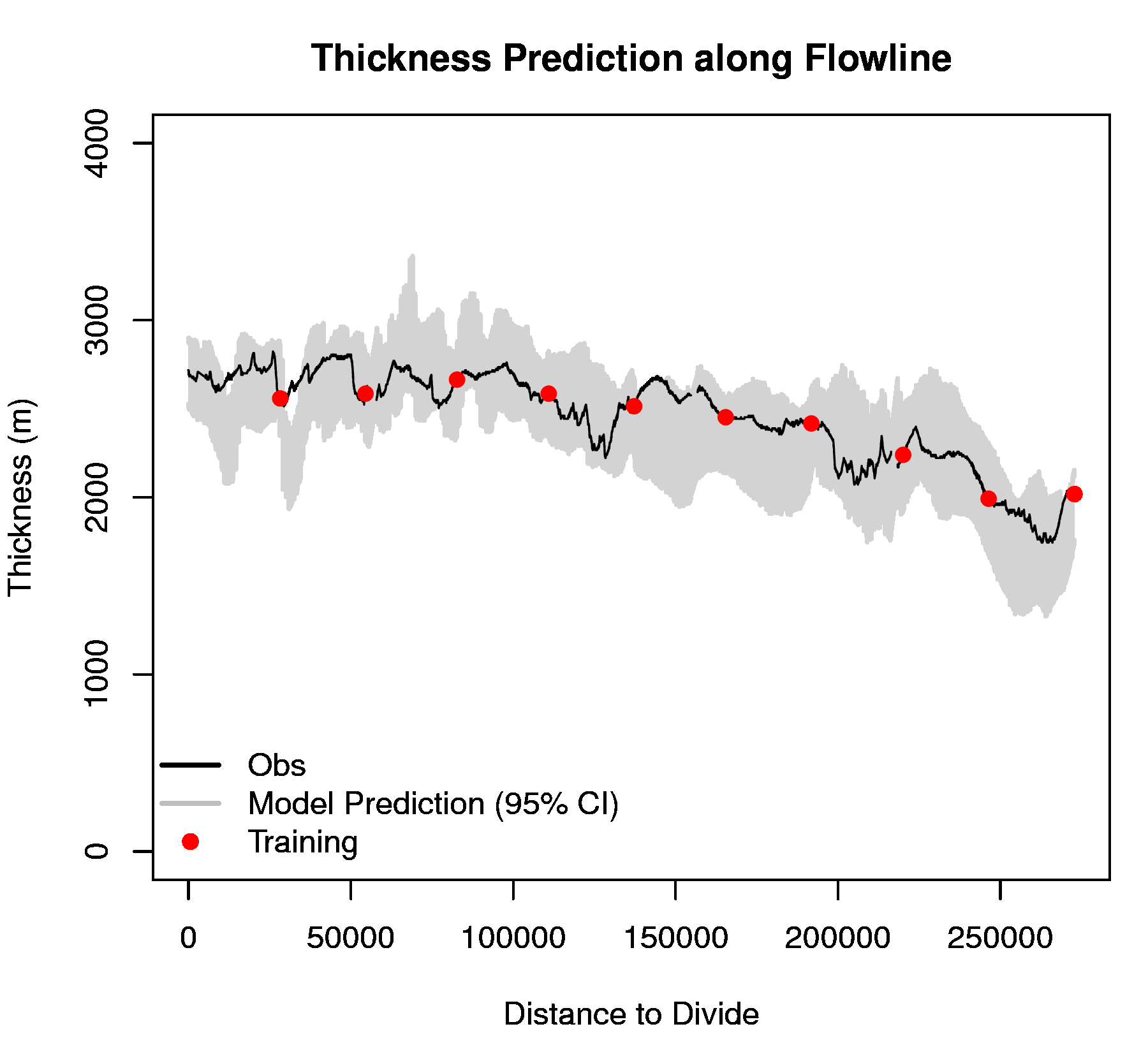}}
		\vspace{0mm}
		\subfloat[]{\includegraphics[width=0.35\textwidth,clip, trim=0cm 0cm 0cm 6cm]{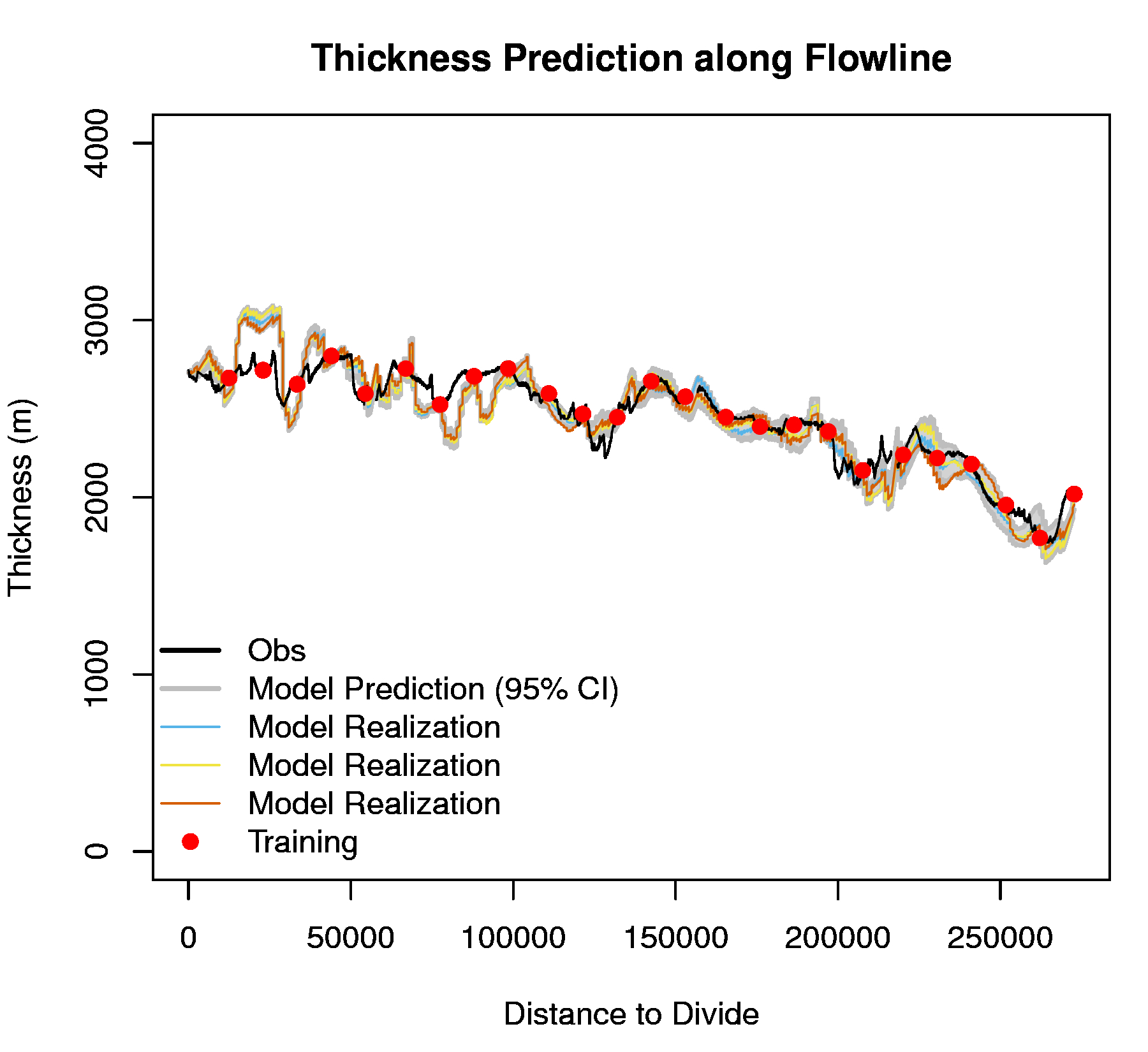}}
		\subfloat[]{\includegraphics[width=0.35\textwidth,clip, trim=0cm 0cm 0cm 6cm]{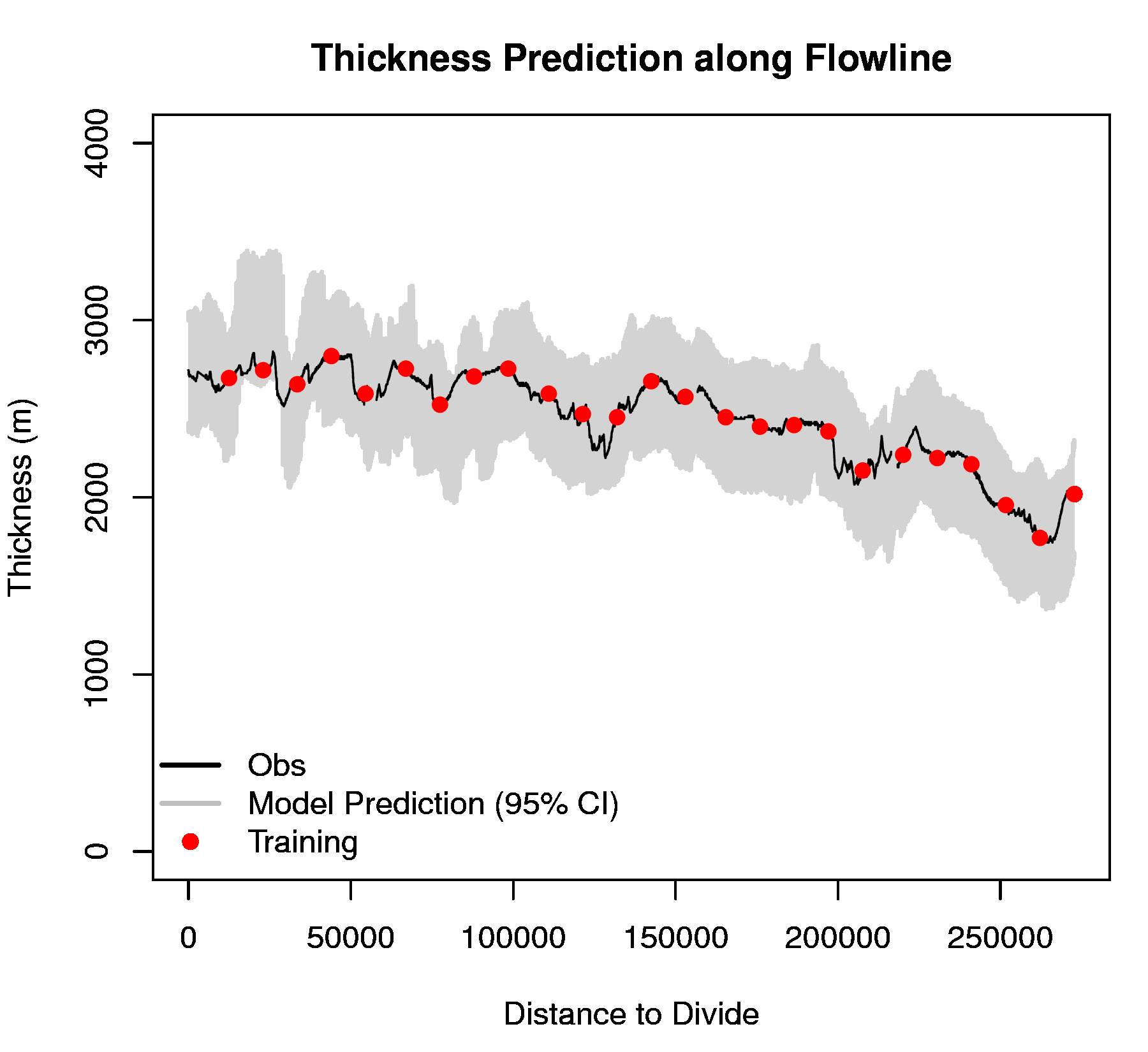}}
		\caption{Predicted ice thickness from the dynamics model using 5 (a), 10 (c), and 25 (e) training locations. The left panel shows the 95\% credible intervals for predicted ice thickness plus observational errors using 5 (b), 10 (d), and 25 (f) training locations.}
		\label{fig:bayes}
	\end{figure}

\begin{figure}[!htbp]
		\centering
		\includegraphics[scale=0.6]{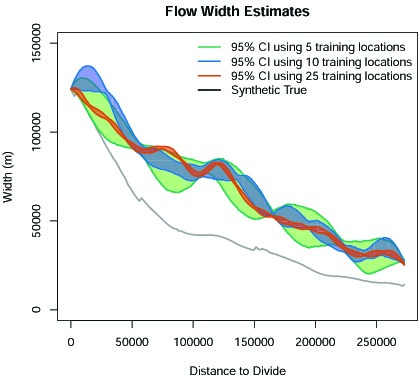}
		\caption{Thwaites Glacier flow width estimates using different number of training locations. More training locations recreates the local feature of the width.}
		\label{fig:bayeswidth}
	\end{figure}

	We now apply the hierarchical approach to the Thwaites Glacier using 5, 10 and 25 training locations. Figure \ref{fig:bayes} shows the model's thickness prediction and Figure \ref{fig:bayeswidth} shows the flow width prediction. Our estimate for the rheologic coefficient is $ 5.085 \times 10^{-17}$($2.259 \times 10^{-18}$, $9.845 \times 10^{-17}$) using 25 training locations; the inference using 5 and 10 training locations are similar, therefore, not reported. As in the simulation study, the uncertainty in thickness prediction increases as its location is further from the divide, reflecting the sensitivity to error in solving ODE.

	\section{Conclusions and future work}\label{sec:5}
	We provide a Bayesian framework with the flexibility of integrating a physical model, multiple observations and uncertainty sources. We show via simulated example and observational data that there are some advantages of our Bayesian approach compared to a non-statistical approach: (1) we obtain uncertainty estimates for the parameters of the physical model, and (2) we provide uncertainties for our predictions of ice sheet thickness. 
	
	There are some important scientific implications of our study. Using 10 training locations along the glacier can reasonably reconstruct the ice structure in some regions. This implies that spatially sparse thickness observations, about 30 km apart can be used, in combination of good quality surface observations to deduce ice thickness. There is a potential to apply our approach to other regions of Antarctic Ice Sheet for estimating ice thickness.

	Although we would like to use as little as 5 thickness observations for reconstructing the ice thickness along the entire glacier, our current result shows that more data is required; whether we can improve ice thickness prediction using less observations remains as an open question. Our method relies on good quality surface observations, and the physical model is very sensitive to small variations in surface elevation; therefore, we pre-smooth the input processes such as the surface elevation and accumulation rate. This pre-smoothing procedure is often performed in practice \citep{pollardinverse,Kamb1986,berliner2008modeling} to eliminate small-scale features that would otherwise result in unreasonable variations. 
	
	Further interesting questions can be studied. For instance, we may be interested in finding out whether the proposed method works well in some regions along the glacier and works poorly in others. An equivalent statement is, to what extent can we use this simple physical law to model complex 3-dimensional ice dynamics? This may allow us to use the simple model for some regions and the complex model for others in combination to reconstruct ice thickness.

	\textbf{Acknowledgments}: The authors would like to thank Knut Christianson and  Nick Holschuh for helpful discussions on the Thwaites Glacier data sets. This work was partially supported by National Science Foundation through (1) NSF-DMS- 1418090 and (2) NSF/OCE/FESD 1202632 and NSF/OPP/ ANT 1341394, (3) Network for Sustainable Climate Risk Management under NSF cooperative agreement GEO1240507, and (4) NSF Statistical Methods in the Atmospheric Sciences Network (Award Nos. 1106862, 1106974, and 1107046). MH and DP were partially supported by (1) and (3), and DP is partially supported by (1), (2) and (3). YG was partially supported by (1) and (4).

	
	\newpage
	\bibliography{Reference}

	\newpage
	\begin{table}[ht]
		\centering
		\caption{Summary of Data Sets}
		\label{tab:data}
		\begin{threeparttable}
			\centering
			\begin{tabular}{|l|l|l|}
				\hline
				Data Set & Spatial resolution \\
				\hline
				Surface velocity, m/year ($V_s$) & 450 m \\
				Surface elevation, m ($E$) & $\sim$ 14 m \\
				Net ice accumulation rate, m ice eq.\tnote{1}/year ($a$) & 55 km \\
				Thinning rate, m ice eq./year ($\tau$)& $\sim$ 1.5 - 5 km\\
				Ice thickness, m ($H$)\tnote{2} & $\sim$ 14 m\\
				\hline
			\end{tabular}
			\begin{tablenotes}
				\item[1]{Ice equivalent (ice eq.)}
				\item[2]{We use only 5, 10 and 25 observed thickness to fit our model to keep our method realistic for applying to other glaciers on WAIS}
			\end{tablenotes}
		\end{threeparttable}
	\end{table}

	\begin{table}[ht]
		\caption{The coverage of 95\% CI's for flow width prediction}
		\label{tab:cov}
		\begin{tabular}{|l|lll|}
			\hline
			& \multicolumn{3}{c|}{Magnitude of Observational Errors }\\
			\hline
			Thickness Observations & Small ($ \sigma_H = 10 $) & Medium ($ \sigma_H = 50 $)& Large ($ \sigma_H = 100 $)\\
			\hline
			n = 5 & 0.962 & 0.960& 0.956\\
			n = 10 & 0.948 & 0.855& 0.770\\
			n = 25 & 0.628 & 0.475& 0.306\\
			\hline
		\end{tabular}
	\end{table}
\end{document}